\documentclass[intlimits,twoside,a4paper]{article}

\usepackage[cp1251]{inputenc}
\usepackage{cmpj3}

\usepackage{bm}

\issue{2021}{24}{3}{33601}
\doinumber{10.5488/CMP.24.33601}

\title[Phase behavior of water in  pores]
{Phase behavior of water-like models in nanoscopic pores of slit shape. 
Predictions from a density functional theory
}

\author[O. Pizio, S. Soko{\l}owski, V. M. Trejos]{
O. Pizio\orcid{0000-0001-8333-4652}\refaddr{label1}\thanks{email: oapizio@gmail.com},
S. Soko{\l}owski\orcid{0000-0003-0580-5214}\refaddr{label2},
V. M. Trejos\refaddr{label3}
}

\addresses{
\addr{label1}
Instituto de Qu\'{i}mica, Universidad Nacional Aut\'{o}noma de M\'{e}xico,
Circuito Exterior, 04510, Cd. de M\'{e}xico, M\'{e}xico 
\addr{label2} 
Department for Theoretical Chemistry,
Maria Curie-Sk{\l}odowska University,  Lublin 20-031, Poland 
\addr{label3}
Instituto de Ciencias B\'asicas e Ingenier\'ia, Universidad Aut\'onoma
del Estado de Hidalgo, Carretera Pachuca-Tulancingo Km. 4.5, Col. Carboneras, C.P. 42184,
Mineral de la Reforma, Hidalgo, M\'exico
}


\date{Received April 9, 2021, in final form May 27, 2021}

\begin{document}
\maketitle

\begin{abstract}
We have explored the phase behavior of a set of water-like models
in slit pores of nanoscopic dimensions. The interaction between water and pore walls 
mimics the graphite surface. A version of density functional method is used as theoretical tools. 
The fluid models are adopted from the work of Clark et al. [Mol. Phys., 2006 \textbf{104}, 3561]. They reproduce the bulk water vapor-liquid coexistence envelope 
adequately. 
Our principal focus is on changes of topology of the phase diagram of confined water 
and establishing trends of behavior of the crossover temperature 
between condensation and evaporation on the 
strength of water-graphite interaction potential. 
Growth of the water film on the pore walls is
illustrated in terms of the density profiles. 
Theoretical results are discussed in context of computer simulation findings 
for water models  in pores.


\printkeywords
\end{abstract}

\section{Introduction}

This manuscript
has been prepared as a tribute to Prof. Yuri Kalyuzhnyi, a distinguished
scientist in the field of statistical physics and theory of liquids on the occasion of his 70th birthday. 
Prof. Kalyuzhnyi has made several important contributions to the statistical theory of associating fluids.
We appreciate his friendship during last decades. On the scientific side, 
we have  benefited from fruitful discussions with him in the mid of
1990s when we just started the development of the theory of inhomogeneous associating fluids 
with late Doug Henderson in Mexico~\cite{doug1,doug2,doug3}. 
This brief stage of our close collaboration, during Yu. Kalyuzhnyi's stay at UNAM in Mexico
resulted in some novel theoretical insights documented in~\cite{yukal1}.

Graphitic materials are of much practical importance in several applications that involve solid-
fluid interfaces (adsorbents, electrodes, solid lubricants).
Thus, efficient and precise elucidation of the surface properties of graphite is
necessary.  Water wettability is one of the most important phenomena characterizing these 
surfaces.
However, from the general perspective, understanding the 
phase behavior of confined associating fluids and mixtures, which differs 
significantly from their bulk counterpart, see e.g.,~\cite{gubbins1} for the review,
is of undoubtful  fundamental and practical importance.

It is difficult to explore the phase behavior of confined fluids by experimental means. 
Concerning the wetting phenomena of water on graphite or on mica, 
experiments have been performed using different methodological 
tools~\cite{Friedman2013,Kozbial2017,Xu1998}.
Commonly, it is accepted that graphite is hydrophobic.
Experimental results usually are given and interpreted in terms
of the contact angle values and of the work of adhesion.
Summarizing insight of experimental measurements of the water-graphite system in this aspect
is given in table~\ref{tableII} of \cite{Kozbial2017}, which provides
the list of methods and data for the contact angle. 

In general terms, the problem of description of phase behavior of water in different pores
using computer simulation techniques is a nontrivial task.
Rather comprehensive computer simulation data for the phase behavior of water in 
cylindrical and slit-like pores of nanoscopic dimensions
and with different water-solid interaction strength were
reported by Brovchenko et al.~\cite{ivan1,ivan2,ivan3,ivan4,ivan5,ivan6} using Gibbs 
ensemble Monte Carlo methodology for ST2, TIP4P, TIP5P and SPC/E water models.
On the other hand, LAMMPS MD method was applied in~\cite{cummings} to describe liquid-vapor
phase coexistence of TIP4P2005 water model in graphite and mica slit-like pores.

The most important theoretical findings in this area of research follow from the
density functional approach for associating fluids. General aspects of this
kind of approaches have been established some time ago~\cite{Segura1997,Yu2002,Huerta1999}. 
A summarizing detailed description of the methodology and quite comprehensive insights 
concerning the results obtained can be found in more recent works~\cite{feng,chapter}. 
For purposes of the present work, we would like to mention that
the surface phase transitions in the context of wetting 
were discussed in~\cite{patryk1} for the one-site associating fluid model.
The models with two and four-associating sites were studied
as well~\cite{millan1,millan2}. However, the only contribution describing the
calculations of the contact angle is \cite{patryk2}, up to our best knowledge.
Usually, theoreticians resort to the Lennard-Jones potential to
describe the non-associating part of the inter-particle interactions.
On the other hand, the square-well model has become overwhelmingly most popular 
with the design of the so-called
statistical associating fluid theory (SAFT), see e.g.,~\cite{gubbins2,gubbins3}
and its application for the modelling of water in \cite{clark}.
The principal idea behind the modelling of Clark et al.~\cite{clark} is to reproduce the 
liquid-vapor (LV) coexistence of water using square-well attraction and 
site-site chemical association
without resorting to electrostatic inter-particle interactions. 

In this work, we present a continuation of the project, started in \cite{trejos1,trejos2}, 
focused on the study of the behavior of water in porous media. However, the principal
issues considered in the present work have not been explored so far.
They are, the surface phase transitions in water-like models confined to wide slit-like
pores and their relation to the wetting of graphite by water.

\section{Model}\label{model} 

\subsection{Generalities}

We consider a  one-component fluid model of associating molecules. 
Each fluid molecule has four associative sites designated by A, B, C and D,
inscribed into a spherical core~\cite{jackson1,jackson2}.
The set of all the sites is denoted by $\Gamma$. 
The pair intermolecular potential between molecules 1 and 2 depends
on the center-to-center distance and orientations,
\begin{equation}
 u(12) = u_{ff}(r_{12}) + \sum_{\alpha \in \Gamma} \sum_{\beta \in \Gamma} 
u_{\alpha\beta}(\mathbf{r}_{\alpha\beta}),
\label{eq1}
\end{equation}
where
$\mathbf{r}_{\alpha\beta}=\mathbf{r}_{12}+\mathbf{d}_{\alpha}(\omega_1)-\mathbf{d}_{\beta}(\omega_2)$
is the vector connecting site $\alpha$ on molecule 1 with site $\beta$ on molecule~2,
 $r_{12}=|\mathbf{r}_{12}|$ is the distance between centers of molecules 1 and 2,
$\mathbf{\omega}_i$ is the orientation of the molecule $i$, $\mathbf{d}_{\alpha}$
is the vector from the molecular center to the site $\alpha$, see also figure~\ref{fig:0} of \cite{jackson1}.
Each of the off-center attraction sites is located at a
distance $d_{s}$ from the particles' center, $d_s = |\mathbf{d}_{\alpha}|$
($\alpha = A, B, C, D$).
In the model in question, only the site-site association  AC,  BC, AD, and BD is allowed,
all association energies are assumed equal. Specifically, the interaction between sites
is given as,
\begin{equation}
\label{eq:asw}
u_{\alpha\beta}(\mathbf{r}_{\alpha\beta})=
\left\{
\begin{array}{ll}
-\varepsilon_{\text{as}}, & {\rm  if \ \ } 0< |\mathbf{r}_{\alpha\beta}| \leqslant r_c ,\\
 0,  & {\rm  if \ \ }   |\mathbf{r}_{\alpha\beta}|   > r_c ,
\end{array} \right.
\end{equation}
where $\varepsilon_{\text{as}}$ is the depth of the association energy well and
$r_c$ is the cut-off range of the associative interaction. The
model was used in several previous works.

The non-associative part of the pair potential, $u_{ff}(r)$, is given as,
\begin{equation}
u_{ff}(r) = u_{\text{hs},ff}(r) +  u_{\text{att},ff}(r),
 \label{eq:sw}
\end{equation}
where $u_{\text{hs},ff}(r)$ and $u_{\text{att},ff}(r)$ are the
hard-sphere (hs) and attractive (att) pair interaction potential, respectively.
The hs term is,
\begin{equation}
u_{\text{hs},ff}(r) =
\left\{
\begin{array}{ll}
\infty, &  {\rm if \ \ } r < \sigma,\\
0,   &  {\rm if \ \ } r \geqslant \sigma,
\end{array}
\right.
\end{equation}
where $\sigma$ is the hs diameter.
The attractive interaction is described by the SW potential,
\begin{equation}\label{uSW}
 u_{\text{att},ff}(r)=
\left\{
\begin{array}{ll}
0, & {\rm if \ \ } r < \sigma,\\
-\varepsilon, & {\rm if \ \ } \sigma \leqslant r < \lambda \sigma,  \\ 
0, & {\rm if \ \ } r \geqslant \lambda \sigma,
\end{array}
\right.
\end{equation}
where $\varepsilon$ and $\lambda$ are the depth and the range of the 
non-associative attraction potential, respectively.

\subsection{Comments on the bulk fluid model and its properties}\label{sec2.2}

In this work our interest is in four water-like model
each with four associating sites designated as W1, W2, W3 and W4 in \cite{clark}.
In order to obtain a set of parameters to describe density-temperature
projection of the phase diagram of liquid water. the following procedure 
within the SAFT scheme is applied. The free energy of the system is written as
a sum of contributions coming from ideal term and residual terms from
monomer-monomer non-associative interaction and site-site association,
\begin{equation}
  \frac{A}{NkT} = \frac{A_{\text{id}}}{NkT} + \frac{A_{\text{mono}}}{NkT} + \frac{A_{\text{as}}}{NkT}.
\end{equation}
The monomer-monomer term is taken from a Barker-Henderson high-temperature
perturbation expansion up to the second order with respect to a hard-sphere reference,
\begin{equation}
 \frac{A_{\text{mono}}}{NkT} = \frac{A_{\text{hs}}}{NkT} + \frac{A_{1}}{NkT} + \frac{A_{2}}{NkT}\,,
\end{equation}
whereas the association contribution comes out from the first-order thermodynamic
theory of Wertheim~\cite{wertheim1,wertheim1a,wertheim2, wertheim2a} in terms of the fraction of 
molecules not bonded at a site $i$, $\chi_i$,
\begin{equation}
  \frac{A_{\text{as}}}{NkT} = \sum_{i=1}^{M} \left[ \ln \chi_i - \frac{1}{2}\left(\chi_i -1\right)\right] .
\end{equation}
All the details concerning these expressions can be found in~\cite{clark,jackson1,jackson2}
and are not given here to avoid unnecessary repetition. In order to obtain
five unknown parameters, $\sigma$, $\varepsilon$, $\lambda$, $\varepsilon_{\text{as}}$
and $r_c$ (all defined in the previous subsection), Clark et al.
used theoretical predictions for the vapor pressure and saturated liquid density
from equations~(\ref{eq1})--(\ref{eq:sw}) to reproduce the experimental vapor-liquid equilibrium data for water.
The fitting procedure is carried out from the triple point temperature
up to the $0.9$ fraction of the bulk critical temperature $T_{c}$. Due to the degeneracy
of the parameters yielding a successful description of the vapor-liquid
equlibrium, four optimum sets are proposed, see table~\ref{tableII}. They can be referred to as
the models with low-dispersion (high hydrogen bonding) values, i.e., W3 and W4,
and the models with high-dispersion (low hydrogen bonding) values, i.e., W1 and W2,
according to the respective weight of the effects coming from different type
of interactions.

\begin{table}[h]
  \centering
   \caption{
   Optimal parameters for water-like model fluids:
   the diameter of particles, $\sigma$,
   the depth and range of the square well non-associative potential, $\varepsilon$
   and  $\lambda$,
   the cut-off distance of the attractive site-site potential, $r_c$,
   and the association energy, $\varepsilon_{\text{as}}$, see \cite{clark}.
   }
    \vspace{3mm}
   \begin{tabular}{cccccc
   }
  \hline \hline
 Model &  $\sigma  $ (nm)& $(\varepsilon/k)$ (K) & $\lambda$&  $r_c $ (nm)&
     \vspace{0.1cm} $(\varepsilon_{\text{as}}/k)$ (K)  \\[0.5ex]
     \hline
W1 & 0.303420 & 250.000 & 1.78890 & 0.210822 & 1400.00 \\
W2 & 0.303326 & 300.433 & 1.71825 & 0.207572 & 1336.95 \\
W3 & 0.307025 & 440.000 & 1.51103 & 0.209218 & 1225.00 \\
W4 & 0.313562 & 590.000 & 1.37669 & 0.215808 & 1000.00 \\
\hline \hline
\end{tabular}
\label{tableII}
\end{table}

Having these parameters available, we employed a simpler version of the theory for the bulk
models with the aim to incorporate the necessary expressions into the density functional
methodology for inhomogeneous associating fluid. In brief, we used the following 
treatment~\cite{trejos1}.  The free energy density (per unit volume), $a =A/V$ is written as,
\begin{equation}
  a = a_{\text{id}} + a_{\text{hs}} + a_{\text{att}} + a_{\text{as}}\,,
\end{equation}
with the following contributions,
\begin{equation}
  a_{\text{id}} = kT\rho \left[\ln(\rho \Lambda^3)-1\right], \quad a_{\text{hs}}=kT\rho \frac{4\eta-3\eta^2}{(1-\eta)^2},
\quad a_{\text{att}}= -4\eta \rho \varepsilon (\lambda^3-1),
\end{equation}
in other words, the hard sphere contribution results from the Carnahan-Starling equation of state and
the attractive contribution follows from the mean-field approximation. The notation is as
follows,  $\rho=N/V$, $\eta=\piup \rho \sigma^3/6$,
$\Lambda$ is the de Broglie thermal wavelength.
Finally, the association term is considered at the level of TPT1~\cite{jackson1,jackson2}, 
\begin{equation}
  a_{\text{as}}= 4kT\rho \left(\ln\chi_A -\frac{1}{2}\chi_A + \frac{1}{2}\right), 
\end{equation}
with,
\begin{equation}
 \chi_A= \left[1+2\rho\chi_A\Delta_{AB}\right]^{-1}, \quad 
\Delta_{AB} = 4 \piup K_{AB} \left[\frac{1-\eta/2}{(1-\eta)^3}\right]\left(\re^{\varepsilon_{\text{as}}/kT}-1\right),
\end{equation}
and where lengthy equation for the site-site bonding volume, $K_{AB}$, is given 
for example by equation~(20) of \cite{trejos1}. The expressions for the chemical potential, $\mu$, 
and pressure, $P$,
involved to study vapor-liquid equilibrium are omitted for the sake of brevity and to avoid
unnecessary repetition,
they are given by equation (22) and (23) of \cite{trejos1}.
The dimensionless input parameters for all the models in questions are given in table~\ref{TableV}.
In the same table, for the sake of convenience of the reader,  we provide the values for 
the critical temperature that results from our version of the theoretical procedure for 
each of the bulk models. Concerning dimensionless units used below, we introduce them
in the following common form, $\rho^*=\rho \sigma^3$, $T^*=kT/\varepsilon$, $\mu^*=\mu/\varepsilon$.

\begin{table}[h]
  \centering
   \caption{
   Dimensionless parameters for the W1, W2, W3 and W4 four-site water-like models:
   the association energy, $\varepsilon_{\text{as}}^*=\varepsilon_{\text{as}}/\varepsilon$,
   the site-site bonding volume, $K^*=K_{AB}/\sigma^3$,
   the cut-off distance of the attractive site-site potential $r_c^*=r_c/\sigma$ ($d_s/\sigma=0.25$).
   Critical temperature of the models, $T^*_c$, is given in the last 
   column. 
   }
   \vspace{3mm}
   \begin{tabular}{ccccccc
   }
  \hline \hline
 Model \ & $\varepsilon_{\text{as}}^*$ & $K^*$ & $r_c^*$ & $T^*_c$ \\
     \hline
W1 & 5.600  & 0.03820 & 0.6948 & 2.718 \\
W2 & 4.4501 & 0.03202 & 0.6843 & 2.193 \\
W3 & 2.7841 & 0.03046 & 0.6814 & 1.330 \\
W4 & 1.6949 & 0.03424 & 0.6882 & 0.862 \\
\hline \hline
\end{tabular}
\label{TableV}
\end{table}

\begin{figure*}[!t]
\begin{center}
\includegraphics[height=6cm,clip]{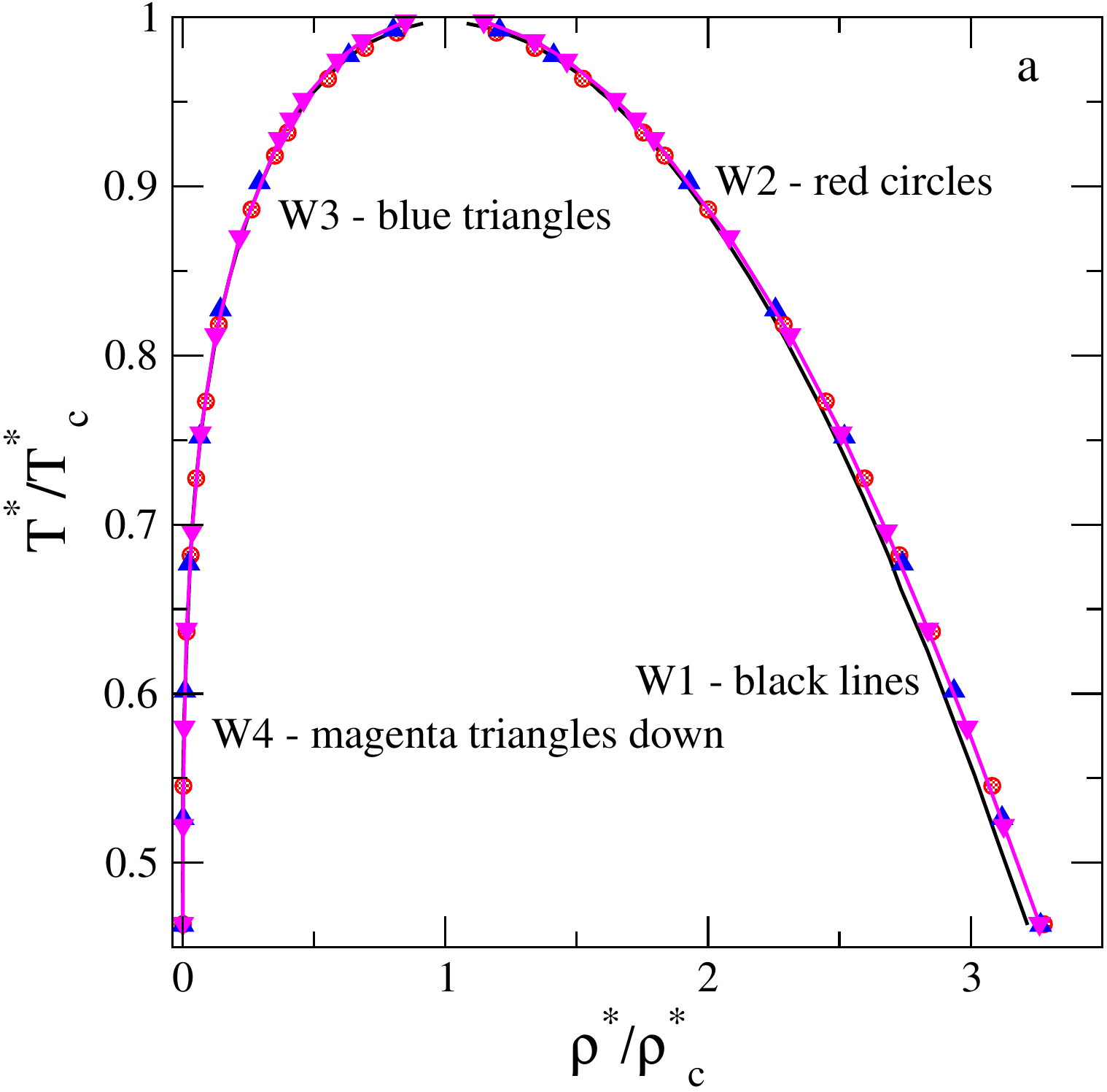}
\includegraphics[height=6cm,clip]{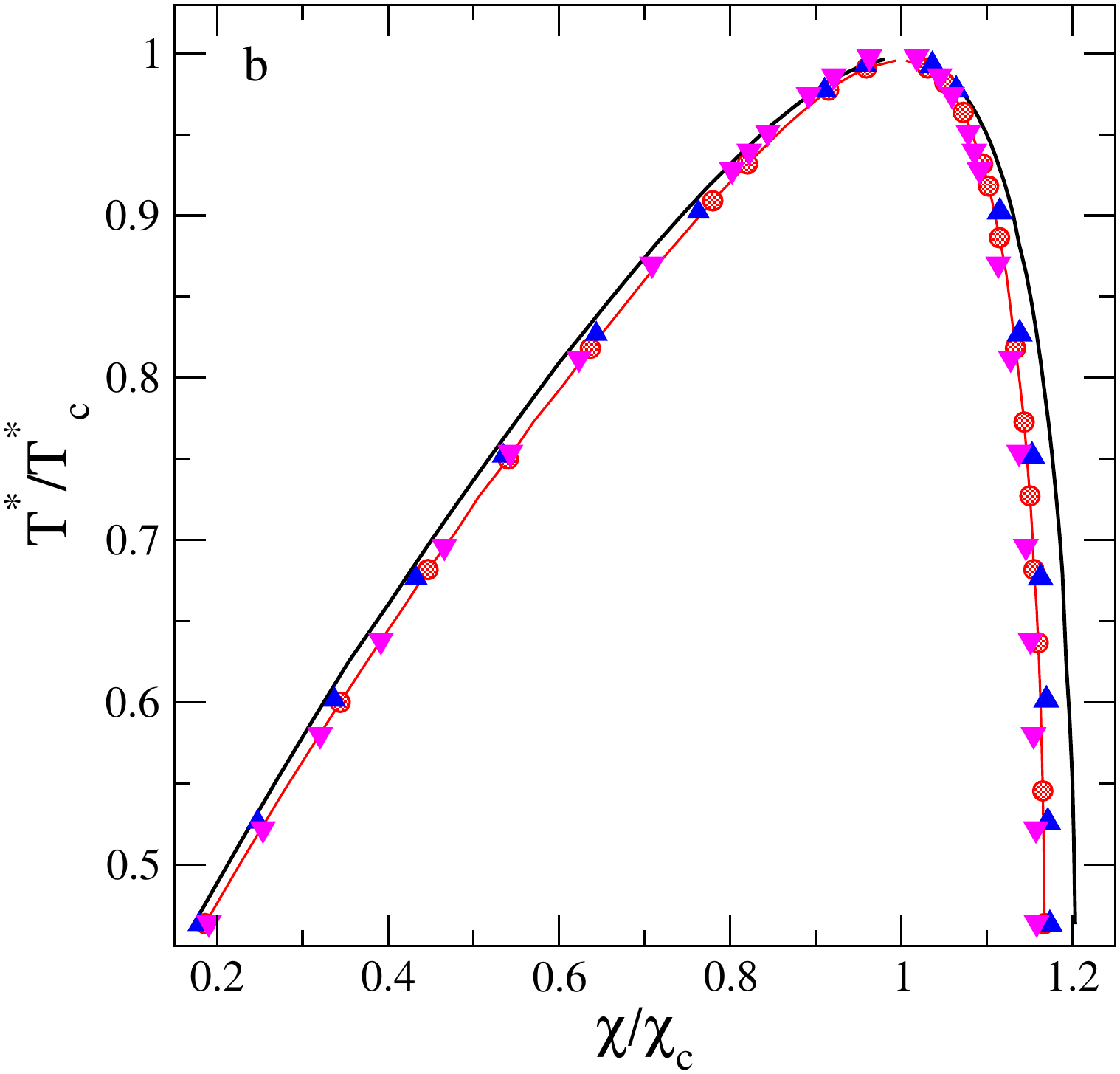}
\end{center}
\caption{(Colour online) The bulk phase diagrams for all models in reduced units. Small deviations
from the law of corresponding states are seen along the liquid branch
of the coexistence envelope. Slightly stronger deviations 
are observed for the fraction of non-bonded species. The nomenclature of lines
and symbols for both panels is given in panel a. 
}
\label{fig:0}
\end{figure*}

\begin{figure*}[!t]
\begin{center}
\includegraphics[height=6cm,clip]{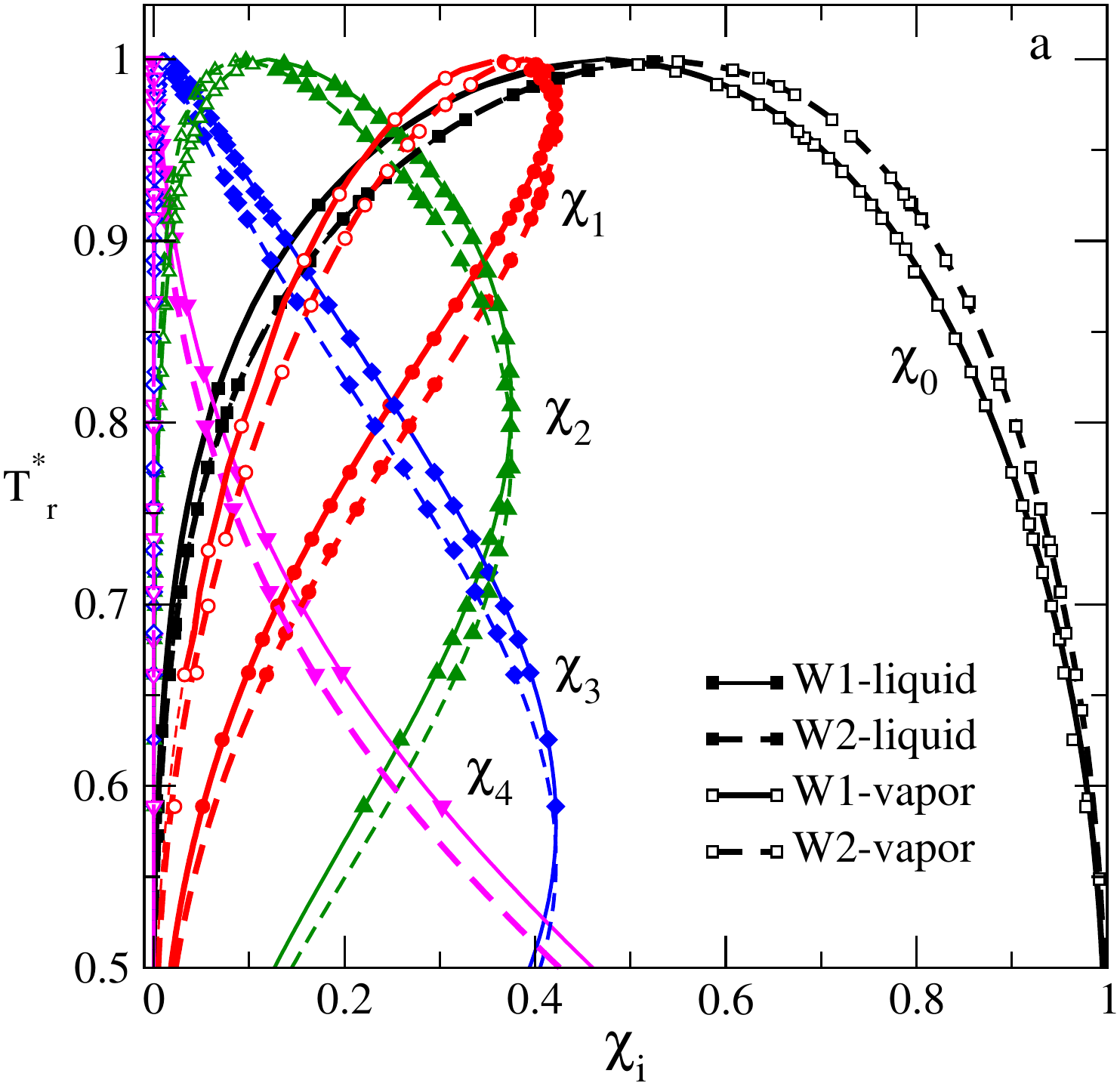}
\includegraphics[height=6cm,clip]{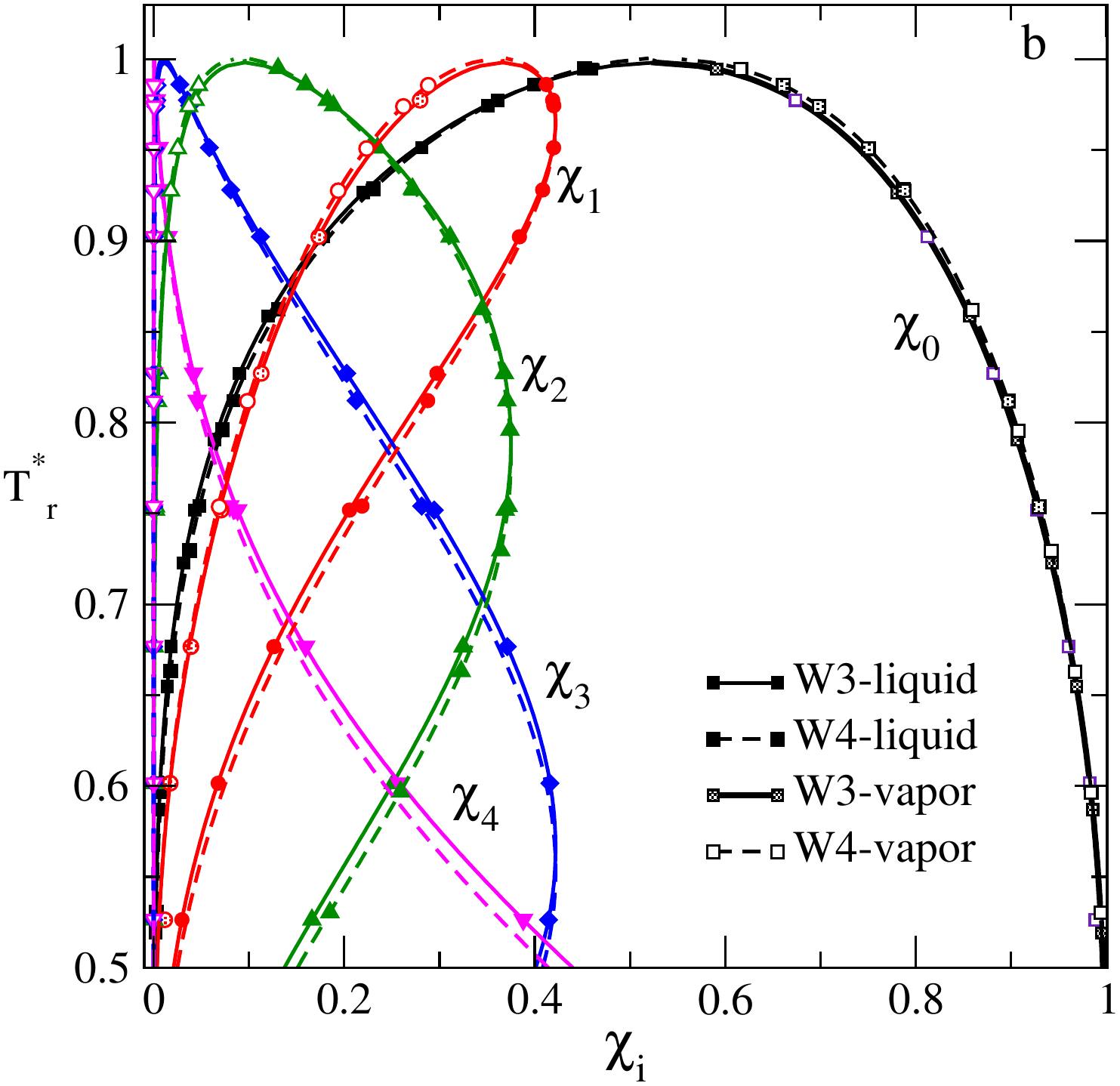}
\end{center}
\caption{(Colour online) The fractions of non-bonded species and
of species participating in different numbers of bonds along
bulk coexistence envelope for W1 and W2 models (panel a).
Panel b: The same as in panel a but for W3 and W4 models.
}
\label{fig:01}
\end{figure*}

Two projections, $\rho^*$ - $T^*$ and $\chi$ - $T^*$, 
of the phase diagrams for the bulk models in question are shown in figure~\ref{fig:0}.
Both are in reduced units, i.e., rescaled to the respective values at a critical point.
Somewhat unexpectedly, one can observe that the models behave according to the
law of corresponding states. Actually, it means that in spite of quite different 
values of parameters for the square-well attraction and for the association energy,
the balance between these two effects is preserved for all four models in question.

Two defficiencies of the applied equation of state should be mentioned. One
of them, mentioned already by Clark et al.~\cite{clark}, is that water anomalies
are not captured by the present modelling. On the other hand, our preliminary
calculations have shown that the critical compressibility factor of this set of models
is not satisfactory. These issues should be analyzed in more detail having in mind the
available analyses, see \cite{kulinskii,ivo}. 

The dimensionless critical temperature for each of the models from table~\ref{TableV} can be converted
into real units to yield: $T_{c}=679.5$~K for W1,  $T_{c}=659$~K for W2; $T_{c}=585.2$~K for W3
and finally $T_{c}=508.6$~K for W4 model. Thus, the performance of the W1 and W2 models in this
aspect is much better compared to other two models (the experimental result is at $647$~K).
The critical temperature for the most popular water models is reported in table~\ref{TableV} of
\cite{vega}. It ranges from 640~K for TIP4P/2005  and 638.6~K for SPC/E models down to 521~K 
for TIP5P water model. In addition to the fraction of nonbonded species at a site,
it is possible to evaluate the fraction of particles in different bonding states~\cite{jackson1,jackson2}.
Previous calculations of these fractions for the model with four associating sites
referred to as associating hard spheres~\cite{holovko,kovalenko}. Here, we would like
to show the fractions of differently bonded species along the bulk coexistence envelope
for all four models in question. 
It appears that in spite of quite different balance between non-associative
attraction and site-site bonding, the behavior of $\chi_i(T^*)$ (on reduced temperature to 
make comparison reasonable) is quite similar. Minor differences are observed between the performance of 
W1 and W2 models, whereas the W3 and W4 models lead to practically indistinguishable results.
The vapor phase is predominantly composed of nonbonded species. In liquid phase, however,
the maximum magnitude of fractions with one, two, three and four bonds is approximately the same.
At a high temperature, the dominant fractions are $\chi_1$ and $\chi_2$, whereas at a low temperature
(of the order of possible triple point), the fluid structure is dominated by $\chi_3$ and $\chi_4$
describing well defined hydrogen bonding network.
From theoretical perspective,
it seems worth to study the application of a more sophisticated procedure to describe 
the non-associative attraction, e.g., the first-order mean spherical approximation
(a limited set of results are reported in~\cite{trejos1}),
and to test how sensitive is the association contribution to the description of the 
reference hard-sphere fluid. These issues require a separate investigation.

\subsection{Fluid-wall interaction}

The pore walls are located at $z=-H/2$ and $z=H/2$.
The external potential, $v(z)$, exerted  on a fluid particle inside the pore
by  walls is,
\begin{equation}\label{Eqv_z}
v_f(z) = v_{fw}(H/2+z)  + v_{fw}(H/2-z) {\ \ \rm for \ \ } -H/2\leqslant z\leqslant H/2.
\end{equation}
The function $v_{fw}(z)$ is given by the Steele's 10-4-3 gas-solid potential \cite{steele,steele2},
\begin{equation}
 v_{fw}(z) =2\piup\rho_s\varepsilon_{\text{sf}}\sigma^{2}_{\text{sf}}\Delta
 \left[ \frac{2}{5} \left( \frac{ \sigma_{fw}}{z}\right)^{10} 
 - \left( \frac{ \sigma_{fw}}{z}\right)^{4} \right. 
\left. - \frac{\sigma_{fw}^4 }
{3 \Delta (z+0.61 \Delta)^3 }
 \right],
\end{equation}
where $\varepsilon_{\text{sf}}$, $\sigma_{\text{sf}}$ are
the cross energy and the size parameters describing solid-fluid interaction. 
The characteristics of graphite are as common, $\Delta=0.335$~nm,
$\rho_s = 114$~nm$^{-3}$, $\varepsilon_{\text{ss}}/k = 28$ K, $\sigma_{\text{ss}}=0.34$ nm.
One frequently used possibility is to evaluate the cross interaction
parameters $\varepsilon_{\text{sf}}$ and $\sigma_{\text{sf}}$
by assuming the Lorentz-Berthelot (LB) combination rules, 
$\varepsilon_{\text{sf}}= \sqrt{\varepsilon_{\text{ss}} \varepsilon}$, and 
$\sigma_{fw}=(\sigma_{\text{ss}}+\sigma)/2$.
However, one can consider these two parameters as free as well. 
In what follows, we use the abbreviation
for the energy of interaction of water-like species
with the pore walls denoted as $\varepsilon^*_{\text{gs}}$,
$\varepsilon^*_{\text{gs}}= 2\piup\rho_s\varepsilon_{\text{sf}}\sigma^{2}_{\text{sf}}\Delta/\varepsilon$.
The pore width is given in dimensionless units as well, $H^*=H/\sigma$.

\section{Theory}\label{theory}

The system is studied using a version of the density functional theory (DF),
described already in detail in   \cite{trejos2,c40,c41}. To avoid unnecessary
repetition, all this stuff is given as appendix.
Rather, we proceed to
the description of the results and novel observations.

\section{Results and discussion: phase diagrams, adsorption isotherm, density profiles}\label{results}

The entire set of water-like models of this study is considered  under confinement in a wide slit-like pore,
$H^*=21$, to describe surface phase transitions and respective phase diagrams.
It is worth mentioning that the computer simulation results of 
Brovchenko et al.~\cite{ivan1,ivan2,ivan3,ivan4,ivan5,ivan6} 
are  obtained for narrower pores, in the case of the slit shape pores the  
maximum width is of the order of $H^* \approx 9.5$. In order to cover
an ample set of effects due to changes of the strength of water-pore walls interaction,
these authors assumed different values for the parameter $\varepsilon_{\text{gs}}$ (ranging from
$-0.39$~kcal/mol to $-4.62$~kcal/mol) to mimic hydrophobic paraffin-like substrate,
moderately hydrophilic, carbon-like surface, and strongly hydrophilic, silica-like surface.  
On the other hand, Srivastava et al.~\cite{cummings} refer to the pores of the maximum width, $H^*=19$,
to describe hydrophilic mica pore and less hydrophilic pore with graphite walls.
It is worth noting that the unique 9-3 water-substrate interaction potential was
assumed in the works from I. Brovchenko laboratory, whereas in \cite{cummings}
the potentials for graphite and mica pores differ from each other.

To keep track with these computer simulation results,
in the present work we employ the following nomenclature of pores: if the value of 
$\varepsilon^*_{\text{gs}}$ parameter is small, so that capillary evaporation is observed
in the entire range of temperature, the pore is termed as hydrophobic. By contrast,
if capillary condensation is observed in the entire temperature range for
big $\varepsilon^*_{\text{gs}}$ values, the pore is termed as hydrophilic. The intermediate
situation, with condensation at high temperatures and evaporation at low temperatures,
is termed as moderately hydrophilic pore. Let us proceed now to the results
for water-like models under different confinement.

\subsection{Hydrophobic pores}

\begin{figure*}[!t]
\begin{center}
\includegraphics[height=5cm,clip]{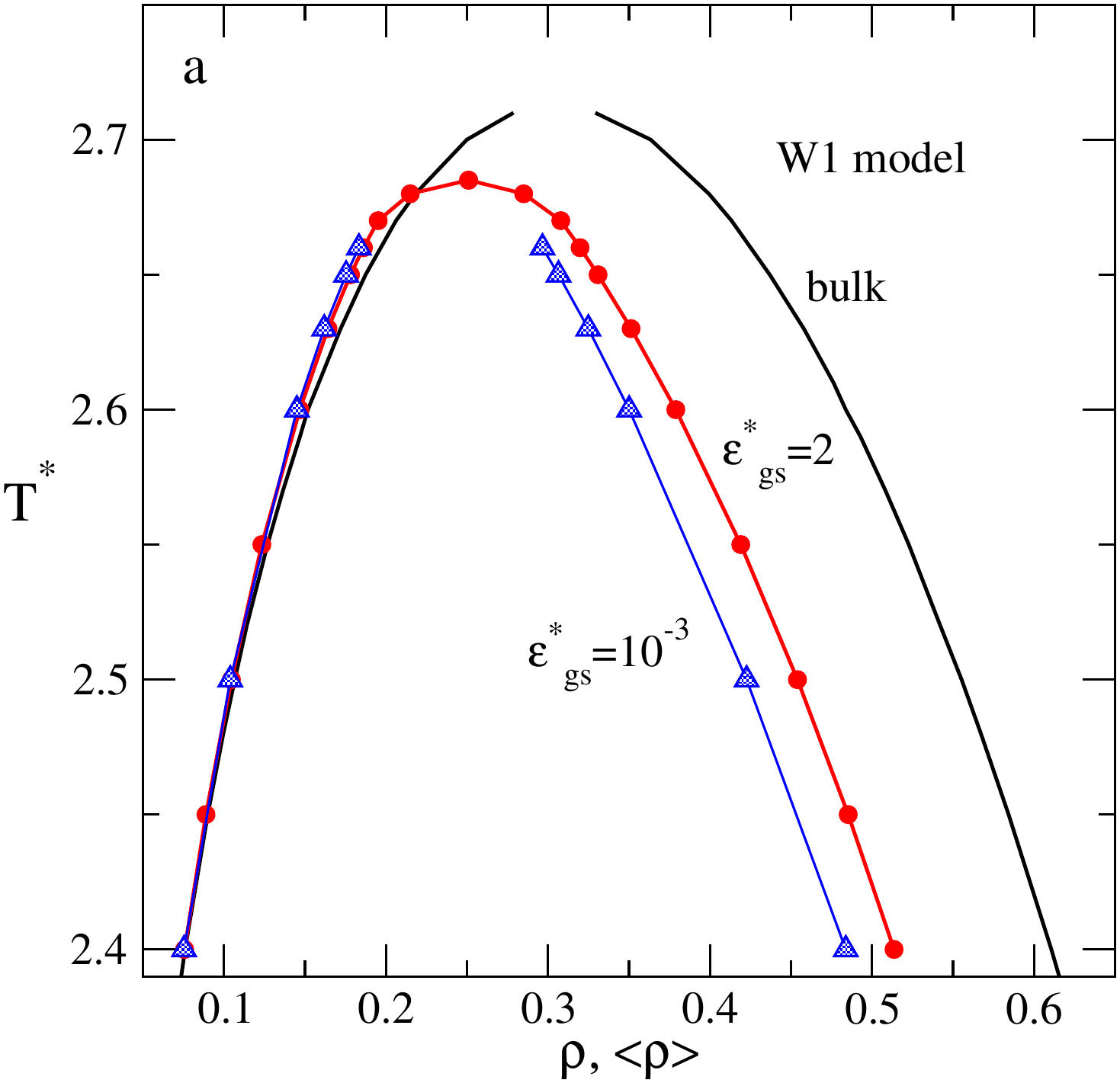}
\includegraphics[height=5cm,clip]{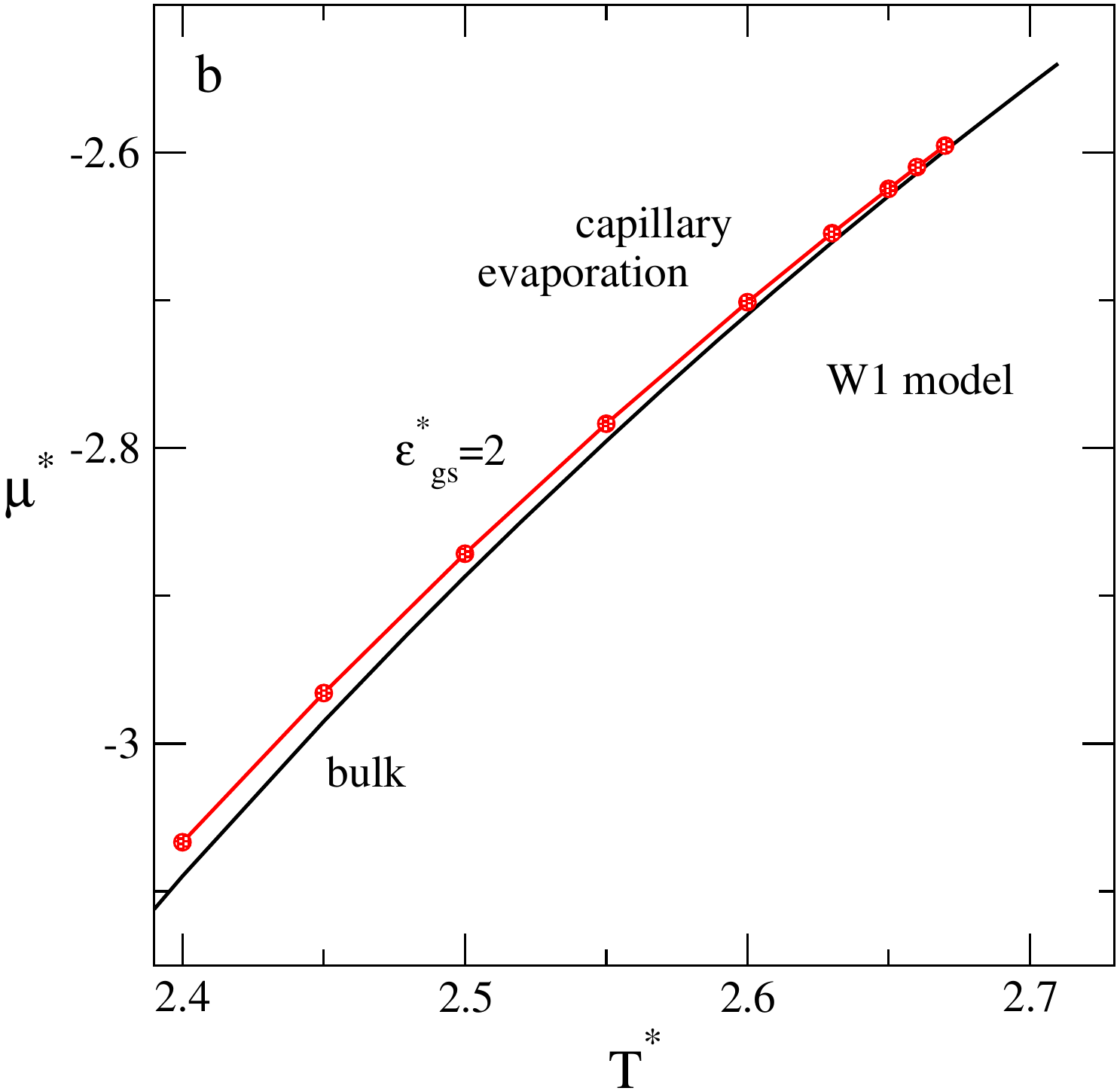}
\includegraphics[height=5cm,clip]{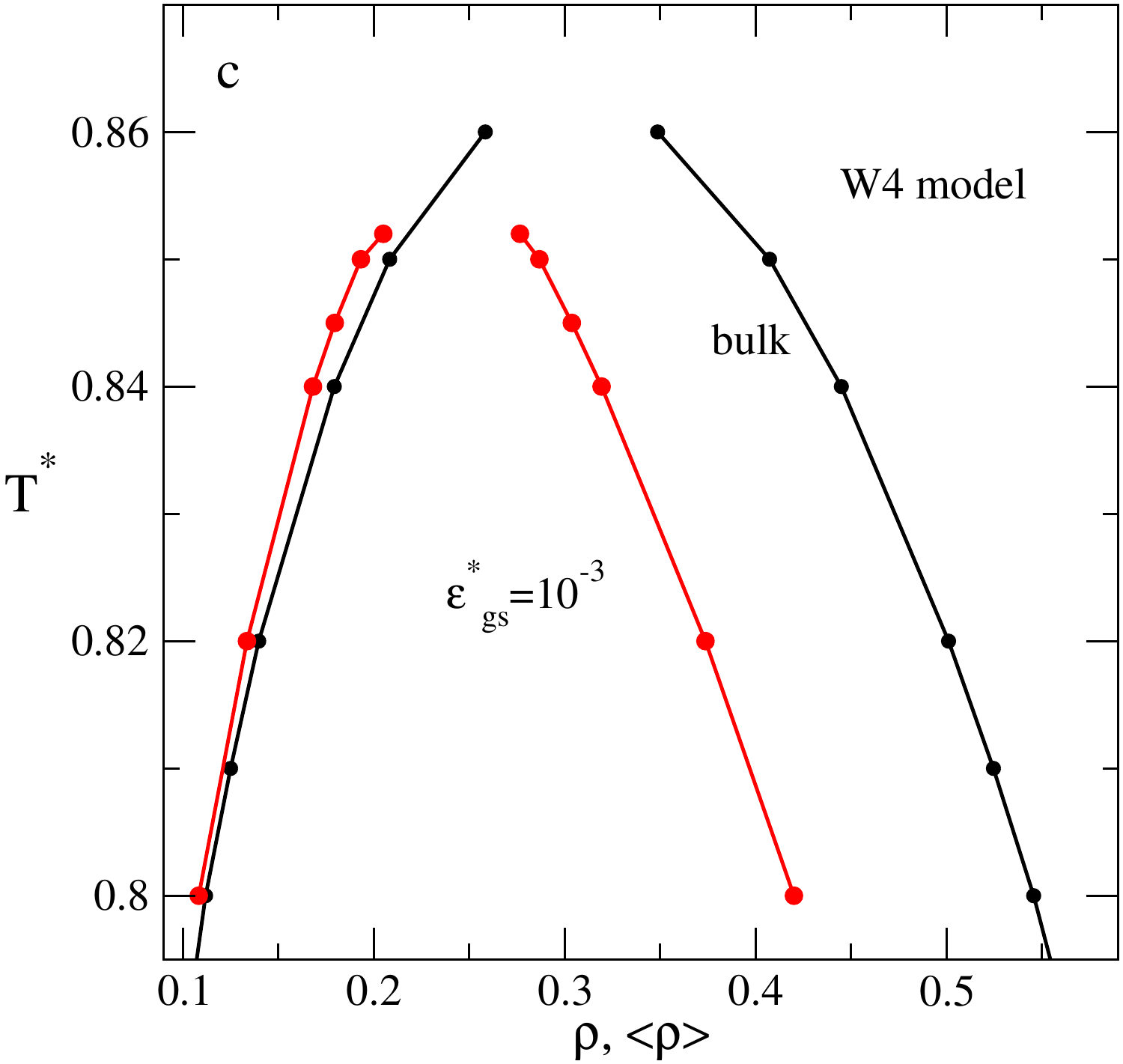}
\includegraphics[height=5cm,clip]{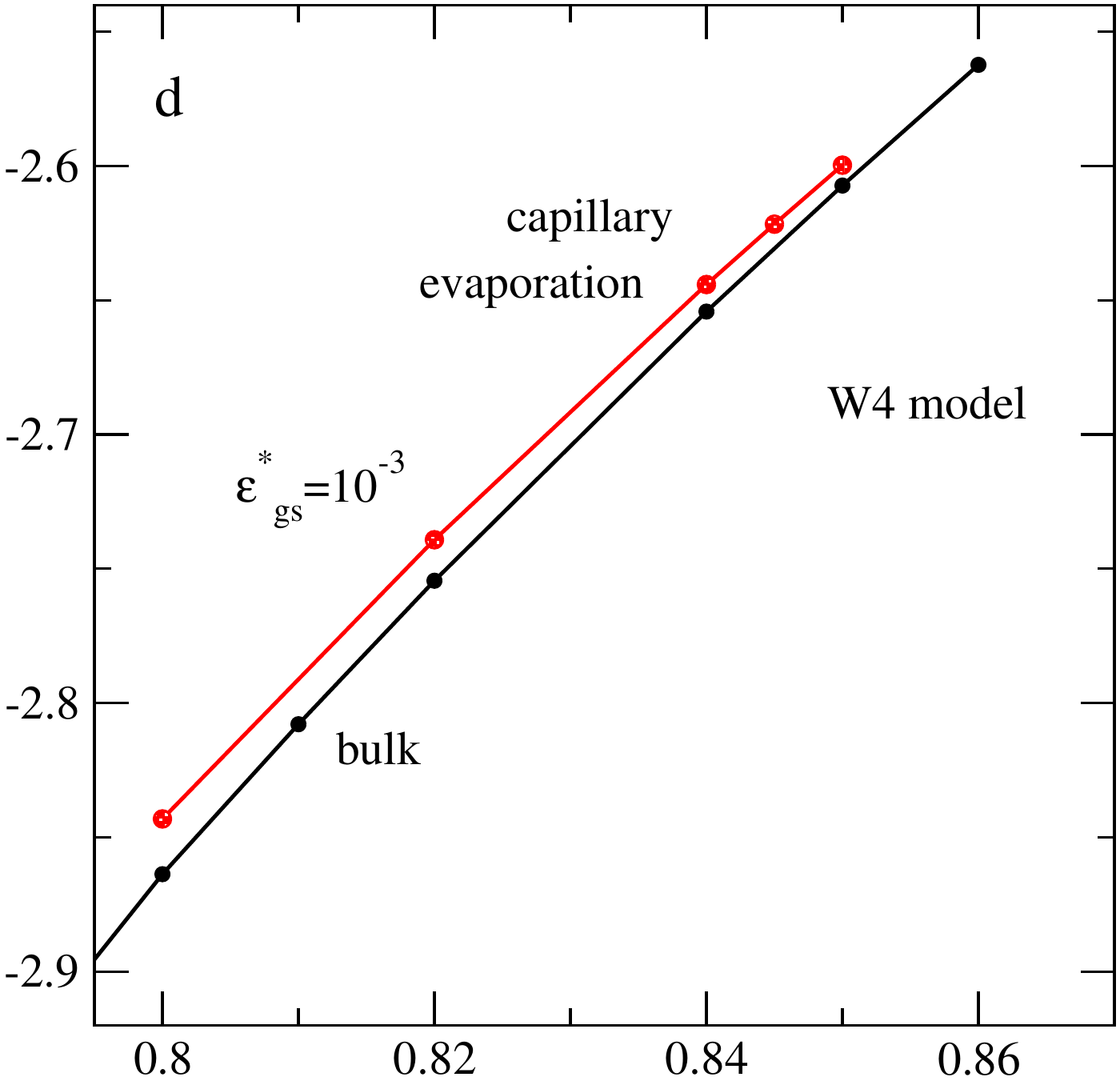}
\end{center}
\caption{
(Colour online) Water-like models in hydrophobic pores.
Panels a and b: The effect of gas-solid interaction energy
on the density-temperature and chemical potential temperature projections
of the phase diagram for W1 model. Panels c and d: the same as in panels a and
b but for W4 model. The pore width is fixed at $H^*=21$ here and in all following figures.
}
\label{fig:1}
\end{figure*}

To begin with, we consider the models W1 and W4 in the hydrophobic environment. 
The first of them, W1 model, is characterized by a rather wide square-well for
nonassociative interparticle attraction, whereas the W4 model has the most narrow
attractive well but the most strong association interaction, cf. table~\ref{tableII}. 
Different projections of the phase diagrams are given in figure~\ref{fig:1}.

Principal theoretical observation is that for both models confined in hydrophobic pore, 
the liquid-density branch of the coexistence envelope pronouncedly departs from its bulk counterpart. 
Magnitude of this effect depends on the value of $\varepsilon^*_{\text{gs}}$ (figure~\ref{fig:1}a).
The vapor branch of the coexistence envelope is much less affected by confinement. The critical
temperature decreases due to confinement and the critical density for both models in the pore
is slightly smaller than the bulk critical density $\rho^*_c$. These DF predictions are in agreement with
the trends observed in computer simulation studies, cf. figure~1 of \cite{ivan6} and
figure~2 of \cite{ivan4}. In the case of a wide graphite-like pore, the critical density is
almost the same as the bulk value, cf. figure~2a in \cite{cummings} but depletion of the 
critical temperature is well marked. The shift of the critical density is correlated with
the depletion of the liquid density branch for confined model w.r.t. its bulk counterpart.

\begin{figure*}[h!]
\begin{center}
\includegraphics[height=6cm,clip]{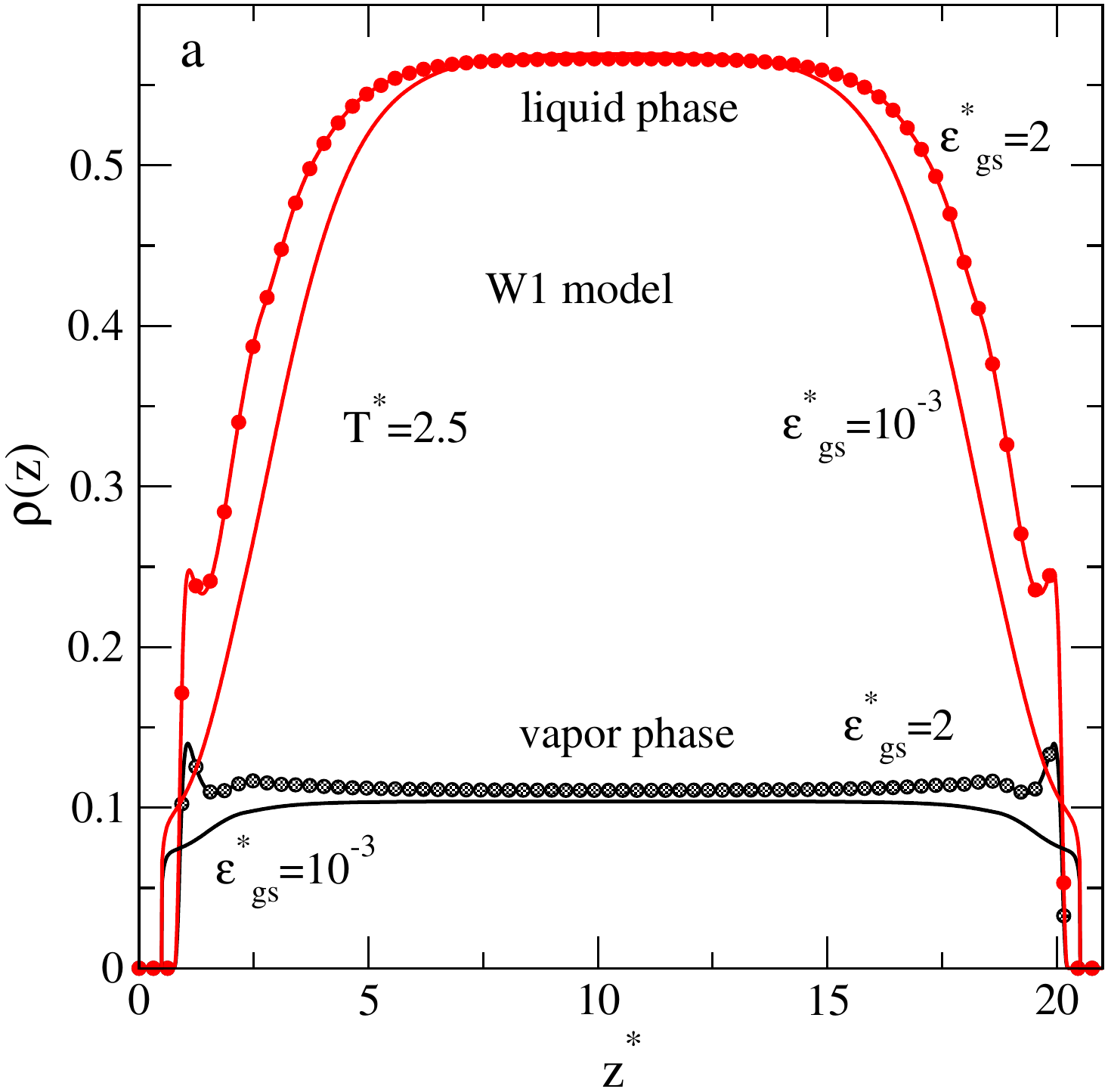}
\includegraphics[height=6cm,clip]{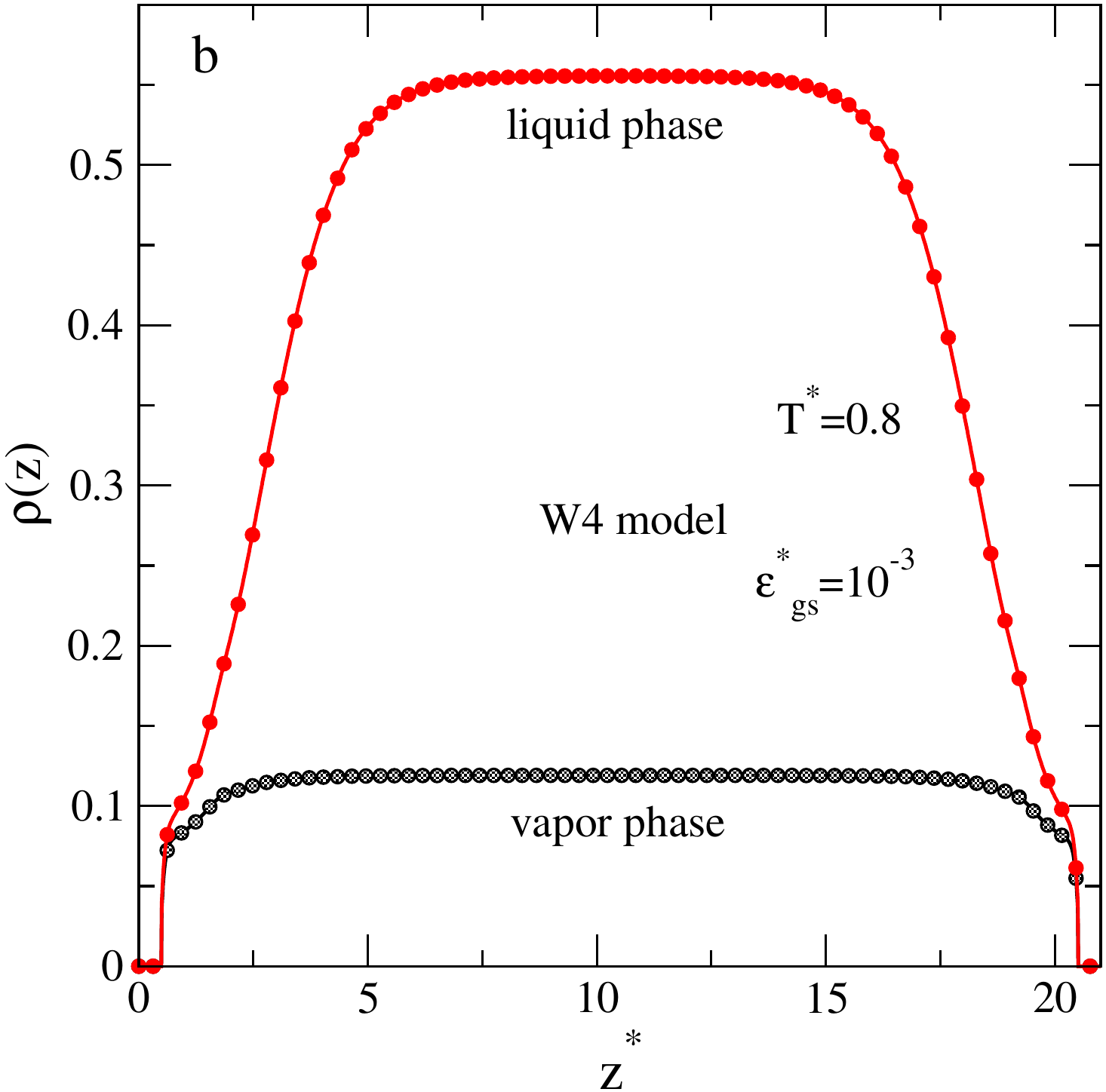}
\end{center}
\caption{
(Colour online) Hydrophobic pores.
Changes of the density profiles of species for the W1
and W4 model upon capillary evaporation transition induced by 
confinement in hydrophobic pore. The reduced tempreature,
$T^*_r=T^*/T^*_c$ is $\approx 0.92$ in both panels.
}
\label{fig:4ab}
\end{figure*}

Changes of water structure upon capillary evaporation in hydrophobic pore are illustrated
in figure~\ref{fig:4ab}. The principal effect is that the fluid density profile is strongly depleted close
to the pore walls. The magnitude of depletion depends on $\varepsilon^*_{\text{gs}}$ and on temperature.
If the density profiles for two models are plotted at the same reduced temperature, 
$T_r^* = T^*/T_c^*$, cf. panels a and b of figure~\ref{fig:4ab} at the same $\varepsilon^*_{\text{gs}}$, the distribution
of molecules in the pore is very similar. In qualitative terms, the shape of the density profiles
coming from density functional theory and from simulation data cited just above is quite similar.

\subsection{Moderately hydrophilic and hydrophilic pore}

The phase behavior of water models in hydrophilic pores is much richer in comparison to hydrophobic pores.
Our first comment refers to the presentation of some plots below.
In several previous works from this laboratory, e.g., \cite{Huerta1999,millan1,trejos2} 
we have used the $\mu_0/\mu$ - $T^*$ plots for the phase diagrams 
[$\mu_0(T^*)$ is the chemical potential of vapor-liquid bulk coexistence at
temperature $T^*$]. On the other hand, the $(\mu-\mu_0)$ versus $T^*$ projection is formally equivalent but it
is more convenient to distinguish between condensation and evaporation in the pore and most importantly 
to establish if the system approaches the wetting borderline or departs 
from it. Moreover, this type of presentation was used in the classical work~\cite{wortis} 
to perform classification of the surface phase transitions based on the strength of attraction
between fluid and substrate. 

\begin{figure*}[!t]
\begin{center}
\includegraphics[height=5.5cm,clip]{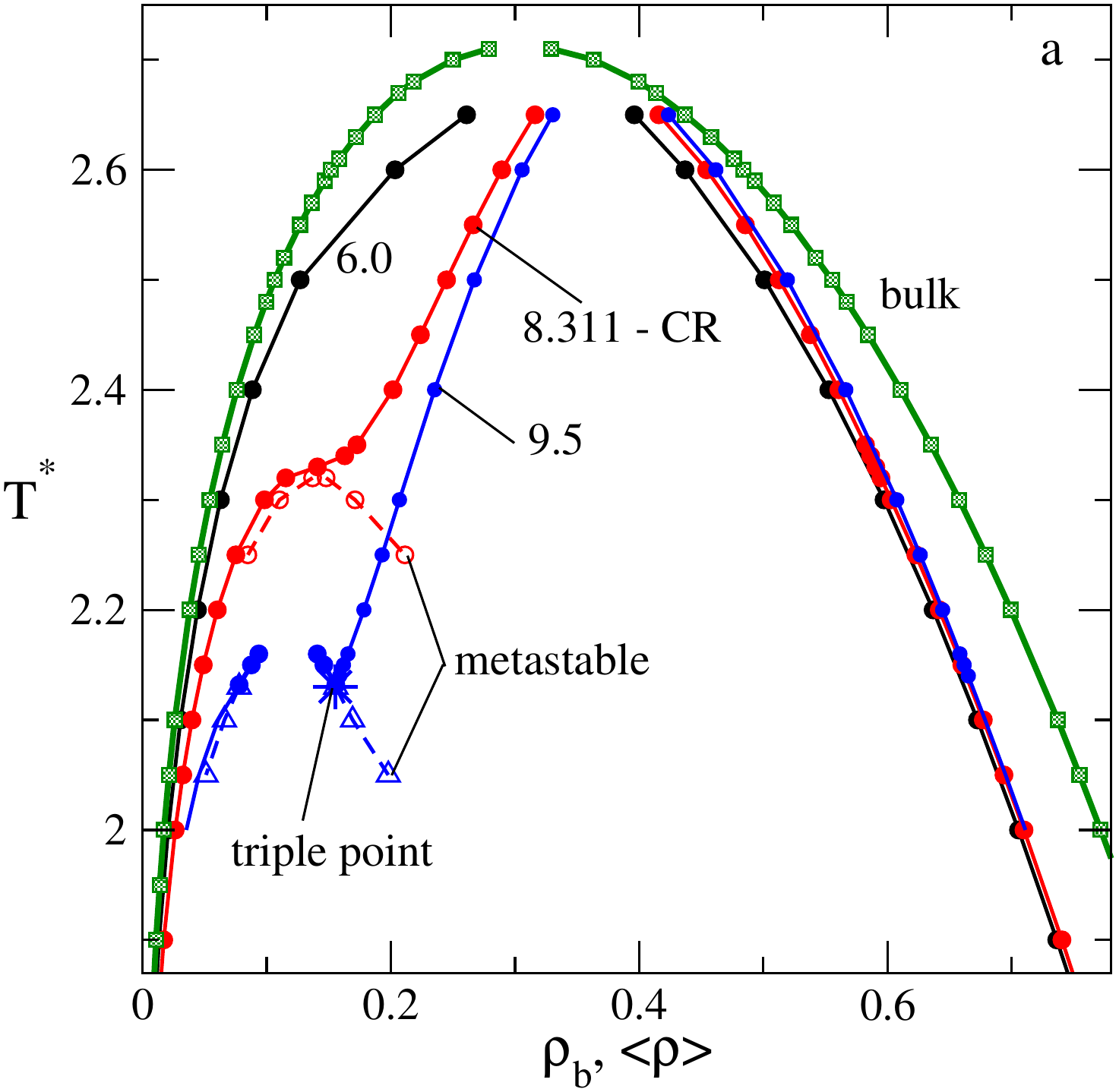}
\includegraphics[height=5.5cm,clip]{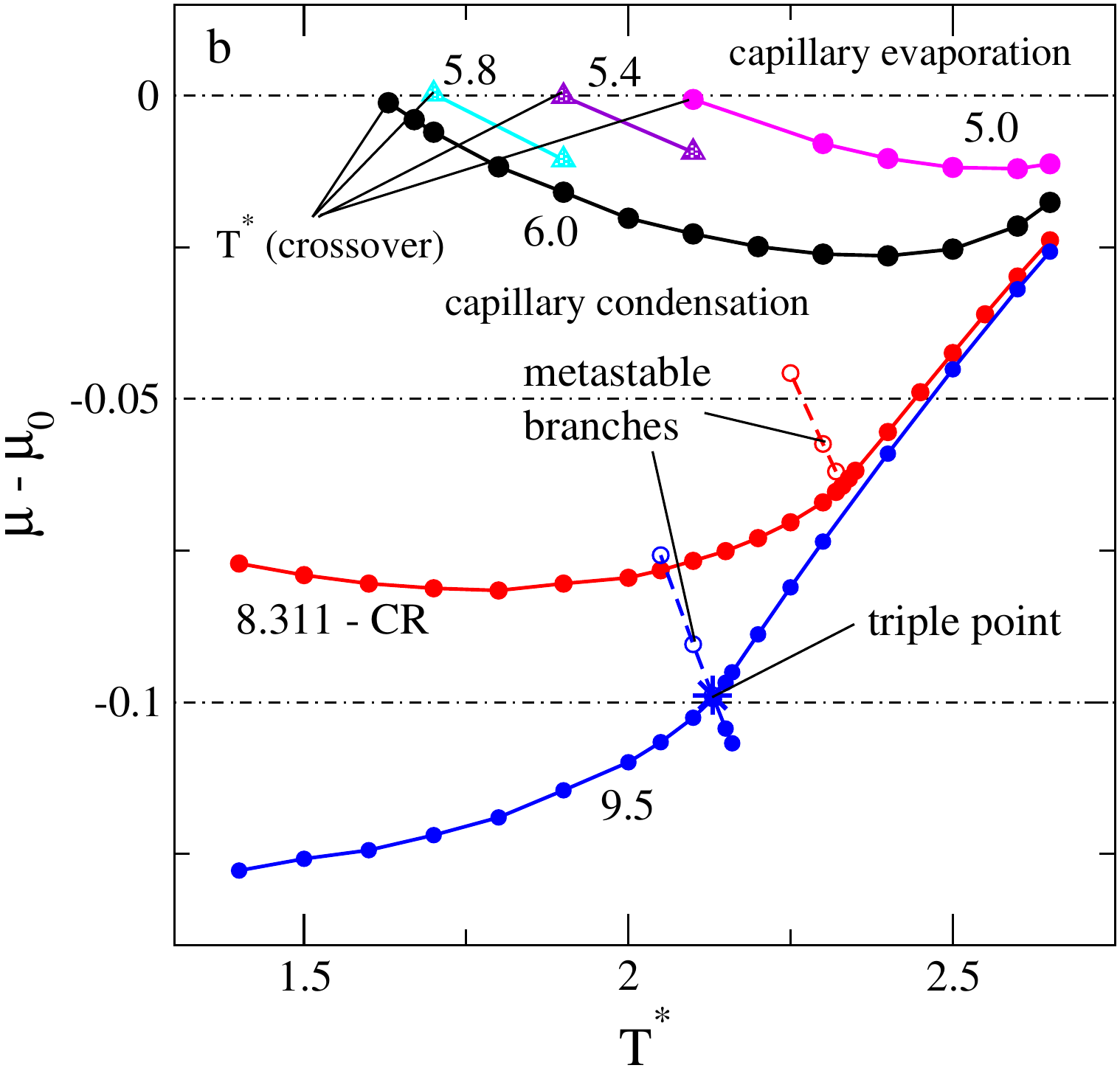}
\includegraphics[height=5.5cm,clip]{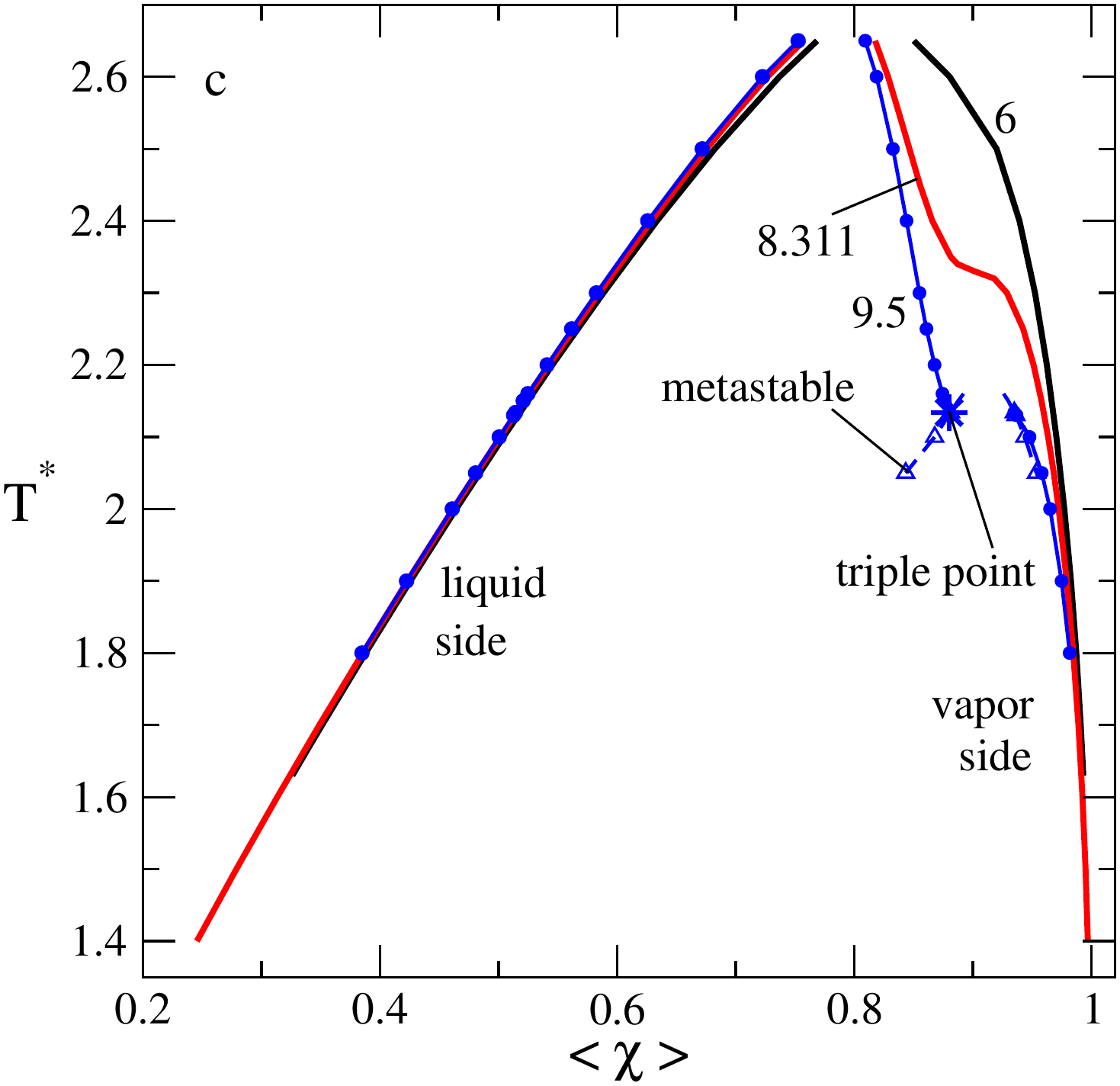}
\end{center}
\caption{(Colour online) Panels a and b: The effect of gas-solid interaction energy on the 
density-temperature and chemical potential-temperature projection of the vapor-liquid coexistence envelope for the  W1 model.
Only fragments of the phase diagrams for intermediate cases, $\varepsilon^*_{\text{gs}}=5.8$
and 5.4 are shown to illustrate crossover between condensation and evaporation regimes.
Panel c: The same as in other panels but in $\left\langle \chi\right\rangle  - T^*$ plane.
The nomenclature of lines and symbols is given in the figures.
}
\label{fig:5ol}
\end{figure*}

The curves in figure~\ref{fig:5ol} illustrate the phase diagrams of W1 model for different values of
the water-substrate attraction in terms of $\varepsilon^*_{\text{gs}}$. 
The case corresponding to 
 $\varepsilon^*_{\text{gs}}=8.311$ that follows from the LB combination rules is marked as CR in the figure. 

In contrast to the discussion concerned with the behavior of water in hydrophobic 
pore, the phase diagrams in figure~5a exhibit shrinking of the coexistence envelope under confinement
that involves 
vapor and liquid branches w.r.t. bulk coexistence. Depression of the critical temperature
is observed. However, the critical density for a confined model now is higher than in the bulk. The
vapor branch can be a single-valued function ($\varepsilon^*_{\text{gs}}=6$) or split due to the 
appearance of additional phase transition ($\varepsilon^*_{\text{gs}}=9.5$).
This is illustrated even better on the chemical potential - temperature projection, figure~\ref{fig:5ol}b.
For moderately hydrophobic pore at relatively low 
values of $\varepsilon^*_{\text{gs}}$, the phase diagram is a single branch that takes negative values along
$\mu-\mu_0$ axis at high temperatures, i.e., the fluid vapor converts into liquid in the entire pore
prior bulk transition (capillary condensation). At low temperatures, the pore filling occurs
at positive values of $\mu-\mu_0$, i.e., capillary evaporation is observed. The crossing point for each
$\varepsilon^*_{\text{gs}}$ value under such circumstances describes transition between two 
regimes. Thus, the coexistence envelope plotted in figure~\ref{fig:5ol}a for the case $\varepsilon^*_{\text{gs}}=6$,
describes condensation in the pore for $T^* > 1.7$ and capillary evaporation at lower temperatures.

If the value of $\varepsilon^*_{\text{gs}}$ is higher, one observes two scenarios. The phase diagram may be a single 
branch, e.g., at $\varepsilon^*_{\text{gs}}=8.311$, or it consists of more branches, 
e.g., $\varepsilon^*_{\text{gs}}=9.5$ (figure~\ref{fig:5ol}b).
In the first case, metastable layering is observed whereas in the second case stable part of the diagram
at low temperatures describes the formation of the fluid layer on the pore walls. This part joins
the principal part of the coexistence envelope at a triple point. 
At still lower temperatures, the formation of such a layer is 
metastable w.r.t. the entire pore filling or capillary condensation. 
Moreover, the inclination of the $\mu -\mu_0$ - $T^*$ projection of the
phase diagram at low temperatures is different (figure~\ref{fig:5ol}b). It 
either  tends to the $\mu=\mu_0$ line
or departs from the bulk coexistence in the entire temperature interval studied.

The results from figures~~\ref{fig:5ol}b  do not permit to get deeper insights into the entire vapor
side behavior of coexistence
and in particular to establish what kind of phase is formed upon layering transition.
To do that, one should return to $\left\langle \rho\right\rangle $ - $T^*$ projection, figure~~\ref{fig:5ol}a. 
First interesting 
observation concerns reasonably high temperatures. At a low value of $\varepsilon^*_{\text{gs}}$, e.g.,
$\varepsilon^*_{\text{gs}}=6$ (black lines), vapor phase is quite ``dilute'', 
whereas at $\varepsilon^*_{\text{gs}}=8.311$, the
``dilute'' phase is approximately twice denser than prior to transition to the entire pore filling. 
Anyway, the water film continuously grows prior to condensation transition.
The transition can be interpreted as thin film (vapor) --- filled pore ($\varepsilon^*_{\text{gs}}=6$) or
thicker film (still vapor) --- filled pore  transition ($\varepsilon^*_{\text{gs}}=8.311$). 
However, at even stronger fluid-substrate attraction, the growth of the
fluid film on the pore walls from the ``underlying vapor''
occurs via discontinuous formation of a separate layer phase
that subsequently converts discontinuously into liquid upon condensation in the pore. 
At even lower temperatures, vapor converts into liquid discontinuously but without an intermediate phase. 
Thus, there exists a certain interval of $\varepsilon^*_{\text{gs}}$ and the corresponding
temperature interval in which a water film grows either continuously or discontinuously prior to filling the
entire pore. 
Such changes upon increasing the chemical potential are accompanied
by changes of the fraction of the average number of non-bonded at a site molecules. This 
kind of projections of the
phase diagram are shown in figure~\ref{fig:5ol}c. On the vapor side, the fraction of non-bonded species is
very high, whereas the liquid phase structure is characterized by a much higher fraction of
species bonded between themselves due to site-site association.

It is worth to comment
our observations in view of computer simulation results. Namely, the situation when the critical
density for the confined fluid becomes higher than for the bulk 
due to an increasing hydrophilicity was observed and discussed 
in~\cite{cummings}, cf. figures~2a and 2b of \cite{cummings}. Nevertheless, a layering
transition was not  observed with their model of fluid-substrate interaction.
Computer simulation evidence of changes of the vapor branch of the coexistence envelope in slit-like
hydrophilic pores was documented in \cite{ivan4} (figure~5) and in \cite{ivan5} (figure 10).
These results refer to the ST2 water model. The authors consider that the formation of an additional
phase is attributed to prewetting. We do not share this opinion completely because a complementary
exploration of one-wall phenomena such as wetting for the models in question is necessary to 
make definite conclusions.

\begin{figure*}[h!]
\begin{center}
\includegraphics[height=7cm,clip]{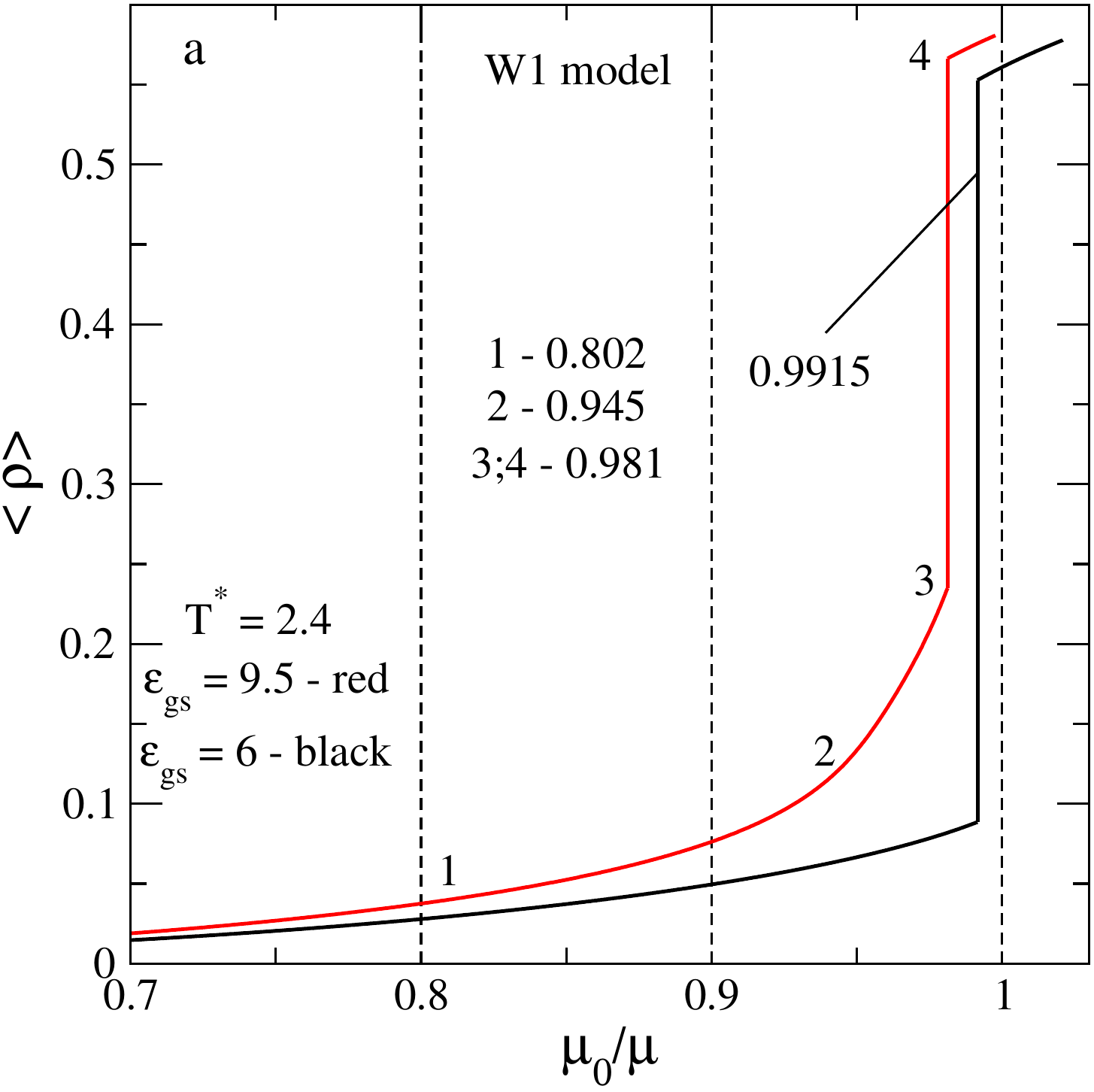}
\includegraphics[height=7cm,clip]{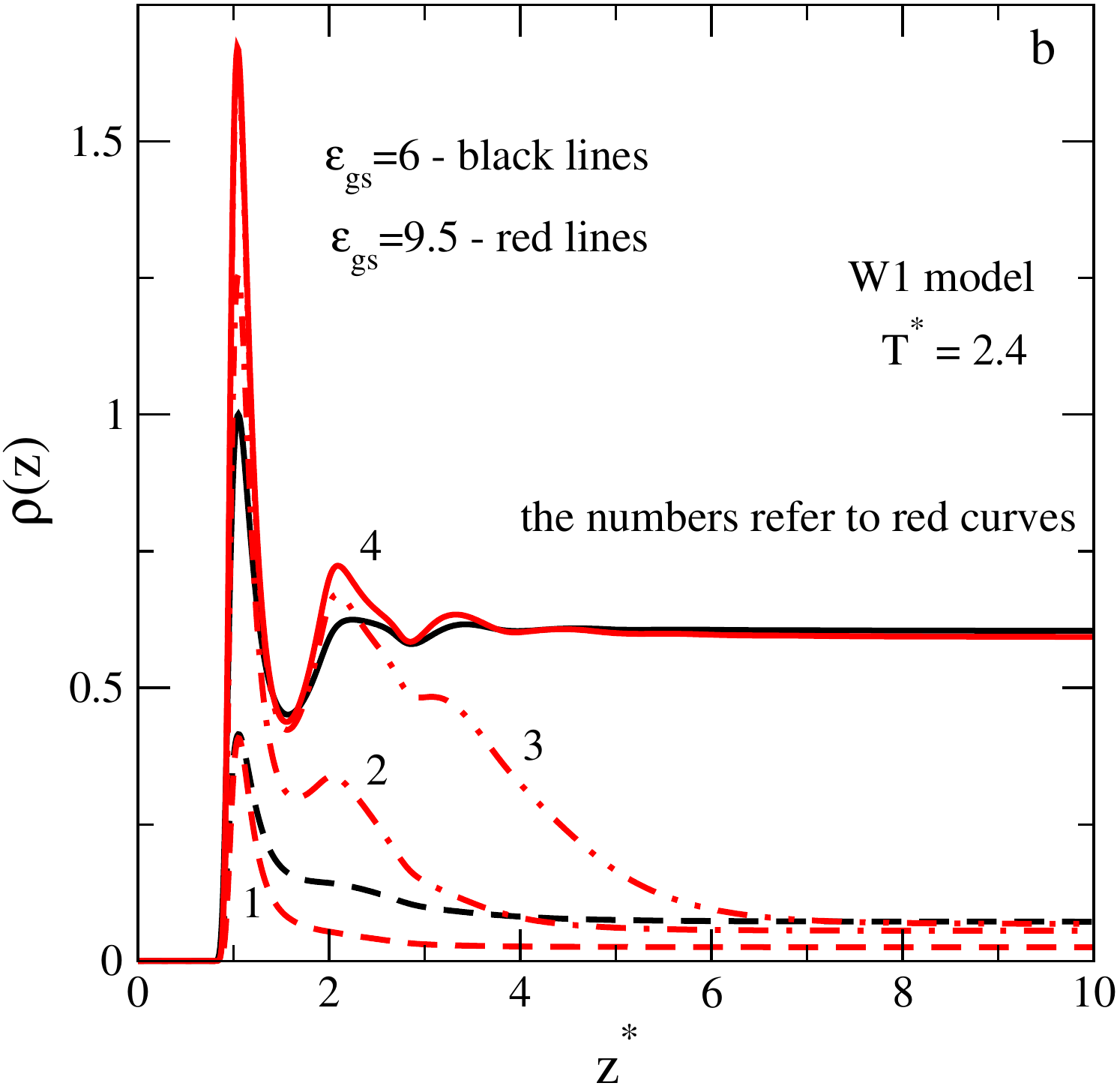}
\end{center}
\caption{(Colour online) Panel a: Adsorption isotherms of the W1 water model on
the substrate with $\varepsilon_{\text{gs}}^* = 6$ (black line)
and $\varepsilon_{\text{gs}}^* = 9.5$ (red line) at $T^*=2.4$.
Panel b: The density profiles of water molecules at states marked on the adsorption
isotherms in panel a. Black lines are for the coexisting densities for the case
of moderately attracting wall, $\varepsilon_{\text{gs}}^* = 6$.
}
\label{fig:6ol}
\end{figure*}

Deeper and complementary insights for the present discussion should come out from the exploration
of adsorption isotherms and fluid structure in the pore.
The phase diagrams in figure~\ref{fig:5ol} were constructed from a large set of adsorption isotherms. 
For purposes of analysis, in figure~\ref{fig:6ol} we show a comparison of adsorption isotherms obtained
at a fixed temperature $T^*=2.4$, and at two values for the strength of water-substrate interaction,
cf. figure~\ref{fig:5ol}a. 

At a higher value of $\varepsilon^*_{\text{gs}}$, $\varepsilon^*_{\text{gs}}=9.5$, the water film on the walls grows faster
and the average density in a wide pore before transition is approximately twice higher than
for the case of $\varepsilon^*_{\text{gs}}=6$. More interesting is the evolution of the density profiles
of molecules shown at certain characteristic points of the isotherm with $\varepsilon^*_{\text{gs}}=9.5$.
First, a layer adjacent to the pore walls is formed (at a point 1). The height of the first maximum
increases if one ``moves'' to point 2. However, the growth of the first layer is accompanied by a
simultaneous formation of the second layer, due to inter-particle attraction (recall that the
square well attraction is rather wide for this model, cf. table~\ref{tableII}) and presumably due to 
inter-particle bonding. At point 3, just before pore filling, both the first and the second layer
are characterized by quite high maxima of the density profile, so that they are almost 
entirely filled. There is a signature of the third layer as well. Upon condensation, the already well formed first 
and second layers do not suffer big changes, in contrast to the third layer. In the entire space,
farther from the pore walls, the vapor drastically converts into liquid.

\begin{figure*}[!t]
\begin{center}
\includegraphics[height=7cm,clip]{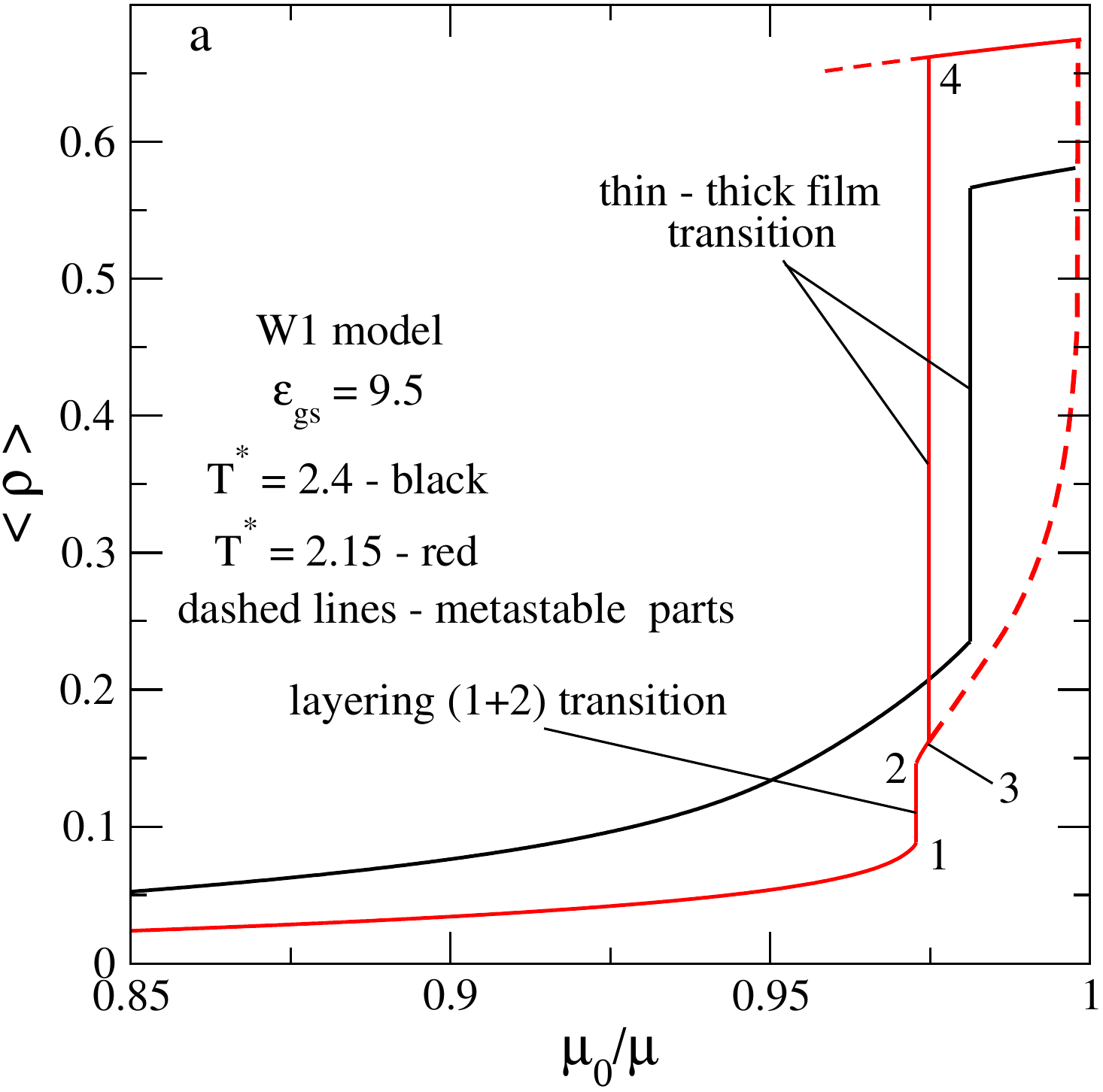}
\includegraphics[height=7cm,clip]{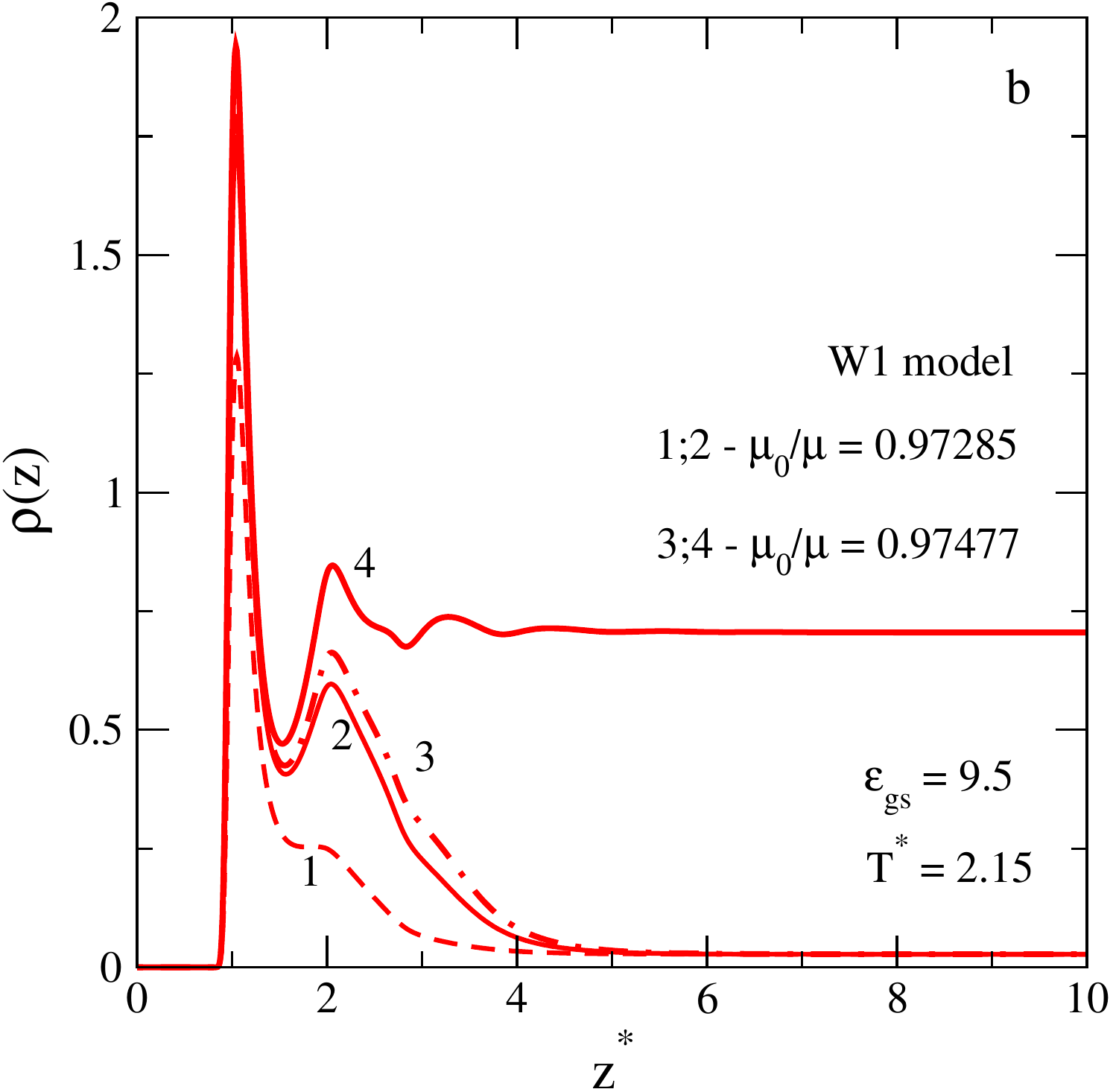}
\end{center}
\caption{(Colour online) Panel a: Adsorption isotherms of the W1 water model on
graphite-like surface $\varepsilon_{\text{gs}}^* = 9.5$  at $T^*=2.4$ and $T^*=2.15$.
Panel b: The density profiles of water molecules at states marked on the adsorption
isotherms in panel a.
}
\label{fig:7ol}
\end{figure*}

In order to elucidate more details of the changes of the structure, we  compared the adsorption
isotherms of the same system with $\varepsilon^*_{\text{gs}}=9.5$, but at two temperatures, figure~\ref{fig:7ol}a. At a 
lower temperature, $T^*=2.15$,  the first-order layering transition occurs (points 1 and 2). At a slightly
higher value of the chemical potential, the pore filling (capillary condensation) takes place. The metastable 
portion of the adsorption isotherm is shown in figure~\ref{fig:7ol}a to illustrate that the growth of 
the film is smooth,
without layer by layer filling. The density profiles in figure~\ref{fig:7ol}b describe how the water structure 
changes along the isotherm. The first layer, adjacent to the wall, is quite dense even before
layering transition. After layering transition, the first layer almost reaches its maximum capacity.
The principal change is the formation of the second layer. There is no signature of the third layer.
This kind of ``bilayer'' structure converts into a liquid structure by completion of the second 
layer together with the filling of the entire pore. 

\begin{figure*}[!t]
\begin{center}
\includegraphics[height=6.5cm,clip]{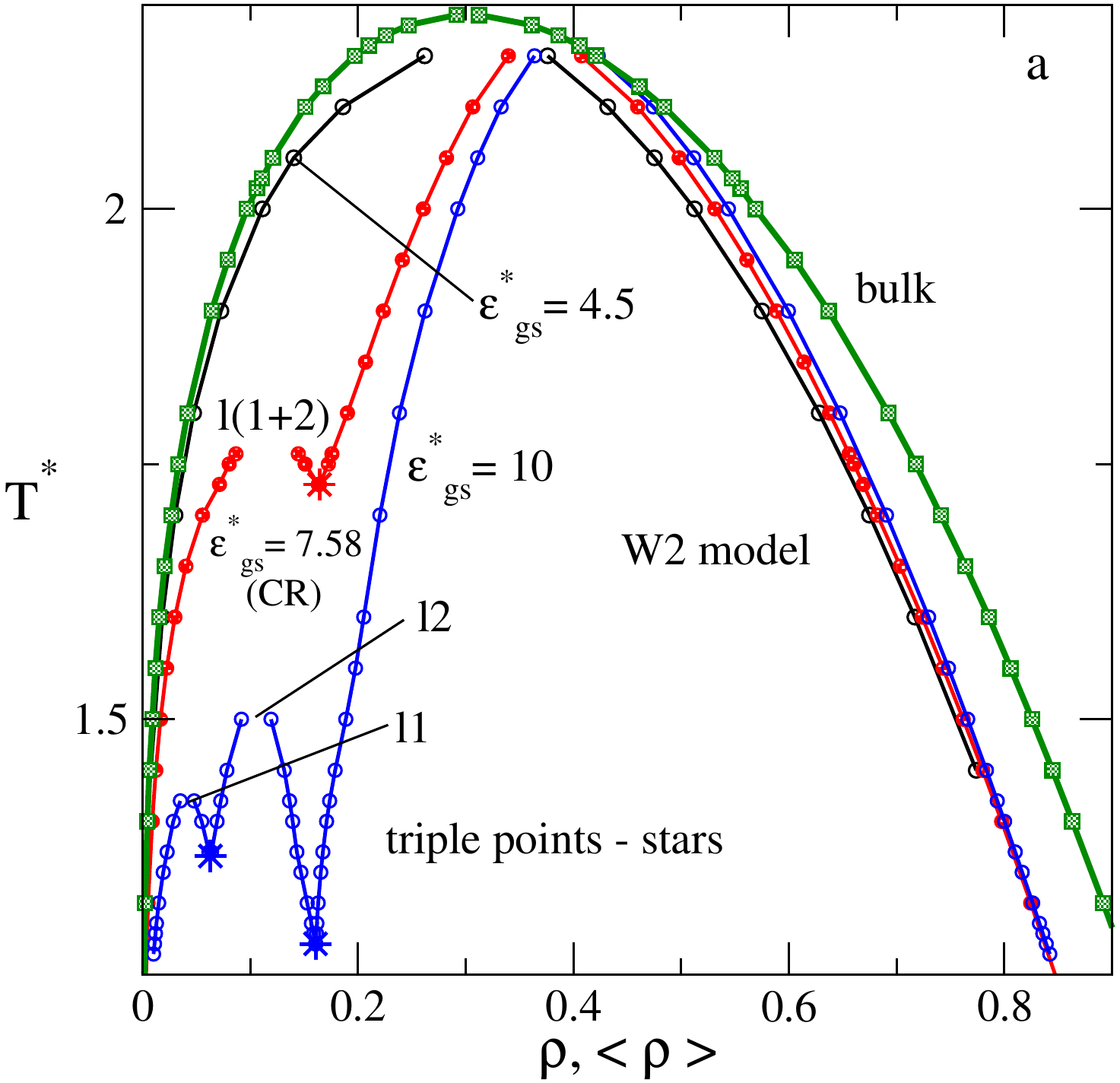}
\includegraphics[height=6.5cm,clip]{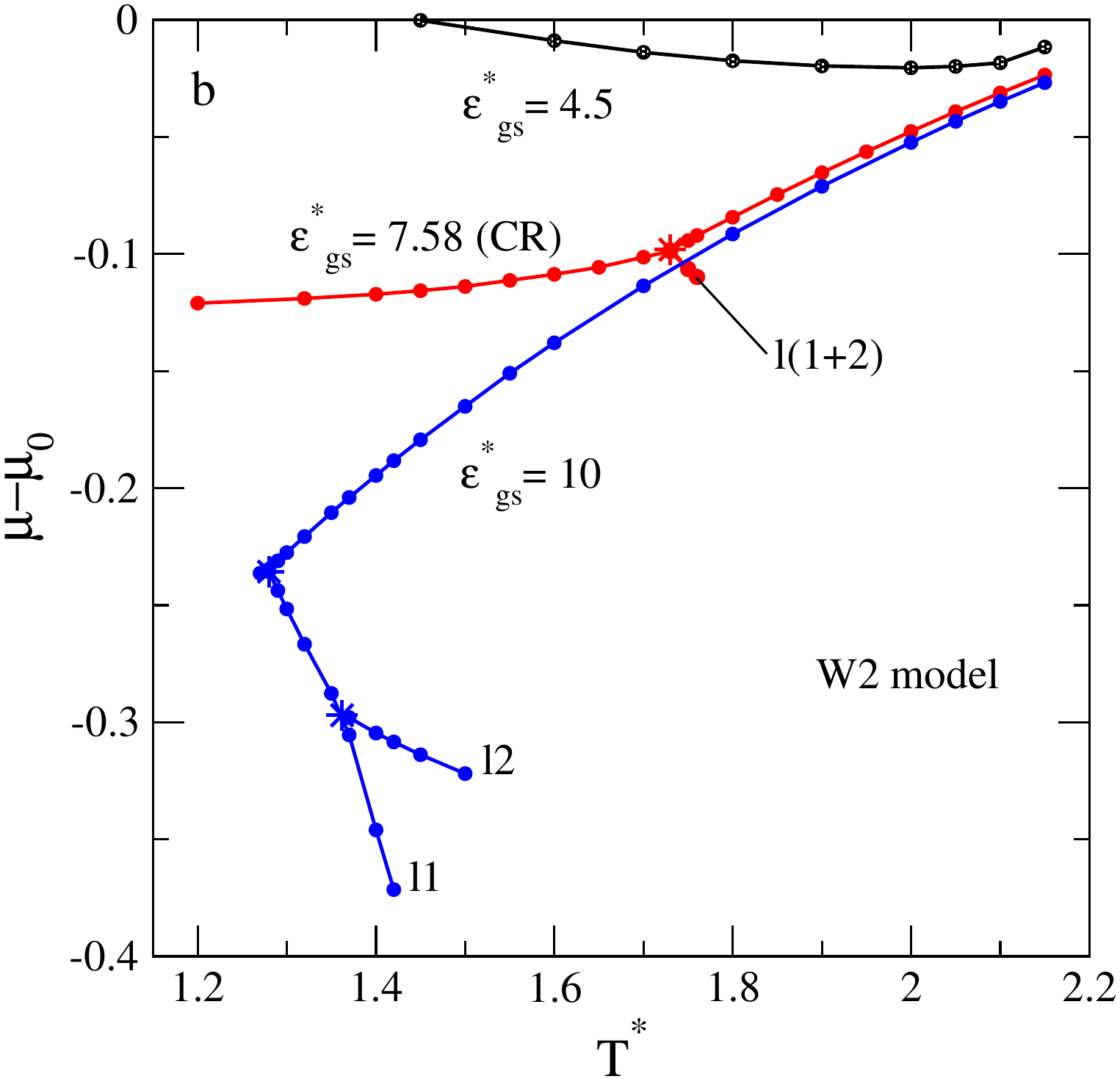}
\end{center}
\caption{
(Colour online) Panels a and b: The effect of gas-solid interaction energy on the 
density-temperature and chemical potential-temperature projections of the vapor-liquid coexistence envelope for the  W2 model
in moderately hydrophilic and hydrophilic pore.
}
\label{fig:8}
\end{figure*}

Proceed now to the description of the results for other water-like models.
According to the parameters from table~\ref{tableII}, one would expect that the phase behavior of the W2 model 
exhibits a similarity to the one for W1 model. This appears to be true. However, some details are
new, in comparison to the findings for the W1. The phase behavior of the W2 model in a wide pore 
at different strengths of water - substrate interaction is shown in figure~\ref{fig:8}. In figure~\ref{fig:8}a, we observe
that if $\varepsilon^*_{\text{gs}}$ is chosen from the LB combination rules, stable layering transition
is observed at certain temperatures, in contrast to W1 with $\varepsilon^*_{\text{gs}}$ chosen according to
combination rules. This transition describes the formation of ``bilayer'' structure. Moreover,
the layering transition envelope joins the principal coexistence at the triple point. Under a further
increase of $\varepsilon^*_{\text{gs}}$, the phase diagram exhibits interesting changes. The layering
 $1+2$ transition in the upper part, decouples into two true, layer by layer, transitions. 
Both of them join at the triple point. The critical temperature of the first layering is lower than of the second one. 
At lower temperature, when the effects of inter-particle and of particle-wall attraction become
stronger, water again forms a film with two layers simultaneously. Therefore, decoupling of the
upper part should be attributed to weakening of inter-particle, presumably non-associative, attraction.
These events are observed in the system when the coexistence line departs quite far from the bulk,
see blue lines in figure~\ref{fig:8}b.

Insights into the changes of water structure close to the pore walls are illustrated in figure~9.
The sequence of changes of the profiles exhibits a similarity to the trends discussed for figure~7b.
However, it is worth mentioning that the first layer, already after the second layering transition 
reaches its maximum capacity. The second layer is also  well filled at these conditions and there 
is no signature of the third layer prior to filling of the entire pore, figure~9a. 

\begin{figure*}[!t]
\begin{center}
\includegraphics[height=6.5cm,clip]{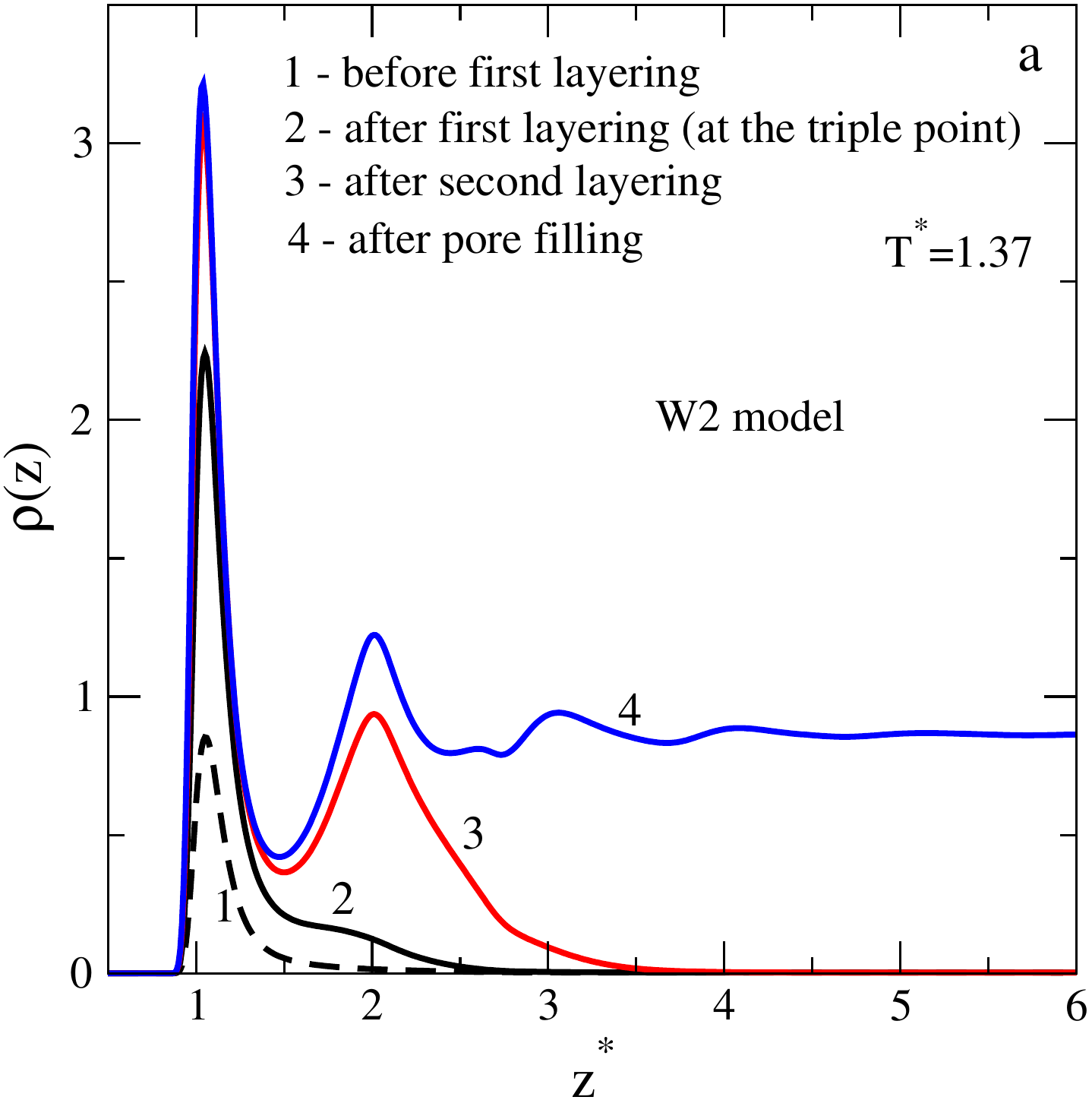}
\includegraphics[height=6.5cm,clip]{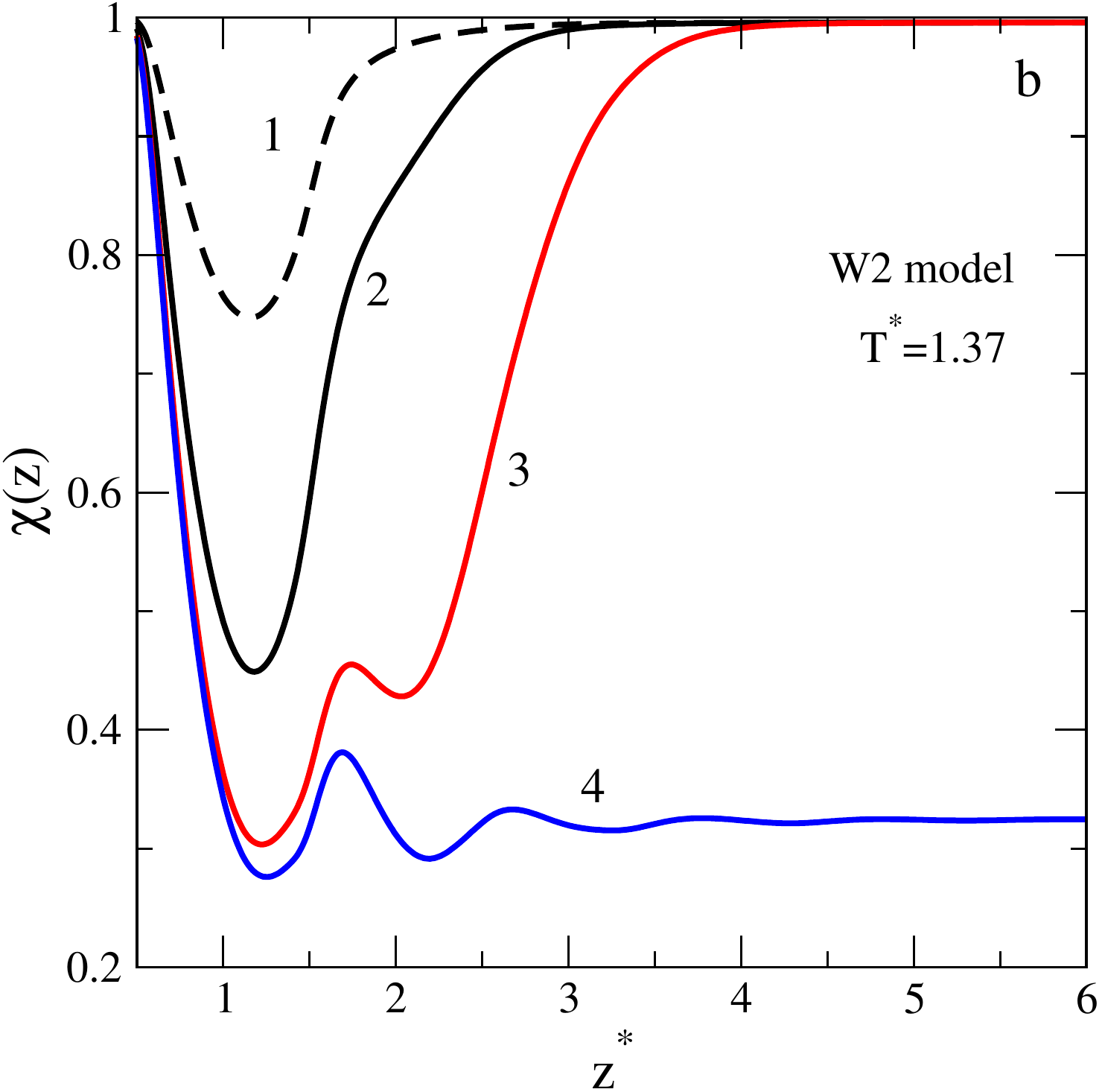}
\end{center}
\caption{(Colour online) Changes of the density profiles of species for the W2 model
at transitions induced by the solid surface.
Panel a: the density profiles before the first layering, at triple point conditions,
after second layering and after pore filling. Panel b: the same as in panel a,
but for the profiles of the fraction of non-bonded species at a site A.
The nomenclature of points is as in panel a.
}
\label{fig:9}
\end{figure*}

Additional view into the
structure is provided in figure~\ref{fig:9}b, where we plot the density profile for particles that are not
bonded at a site. The fraction $\chi(z)$ follows the trends of the behavior of density in figure~\ref{fig:9}a.
Higher local density results in  lower local values for the fraction of non-bonded species or
equivalently causes a stronger bonding tendency. In this plot we observe that after the second layering
transition, the bonding state within the first layer very closely resembles the bonding state of the
liquid phase filling the pore. The second layer is characterized by a high degree of bonding as well.
How good are these values in particular states would require performing quite elaborate
computer simulations.

\begin{figure*}[!t]
\begin{center}
\includegraphics[height=6.5cm,clip]{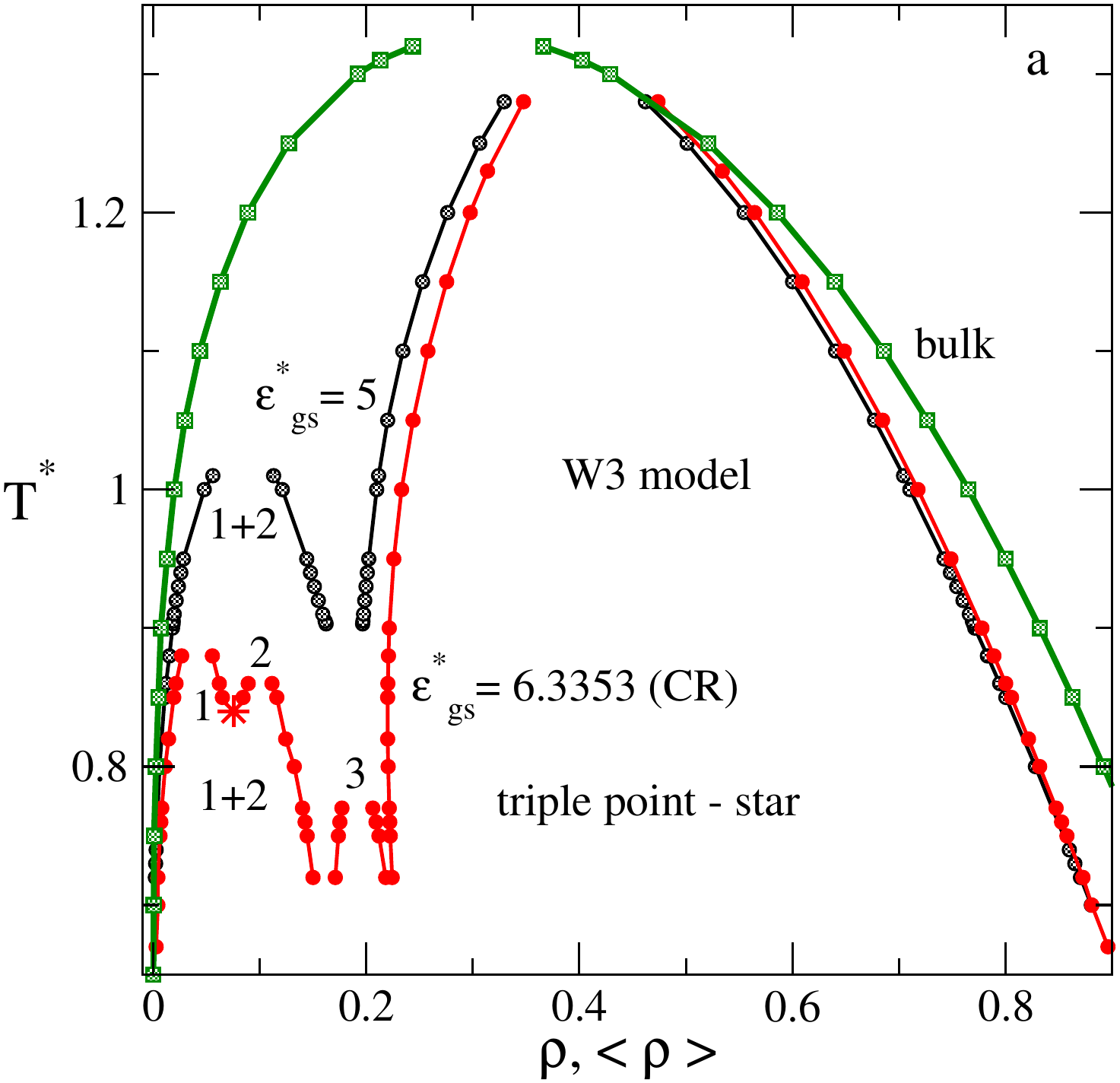}
\includegraphics[height=6.5cm,clip]{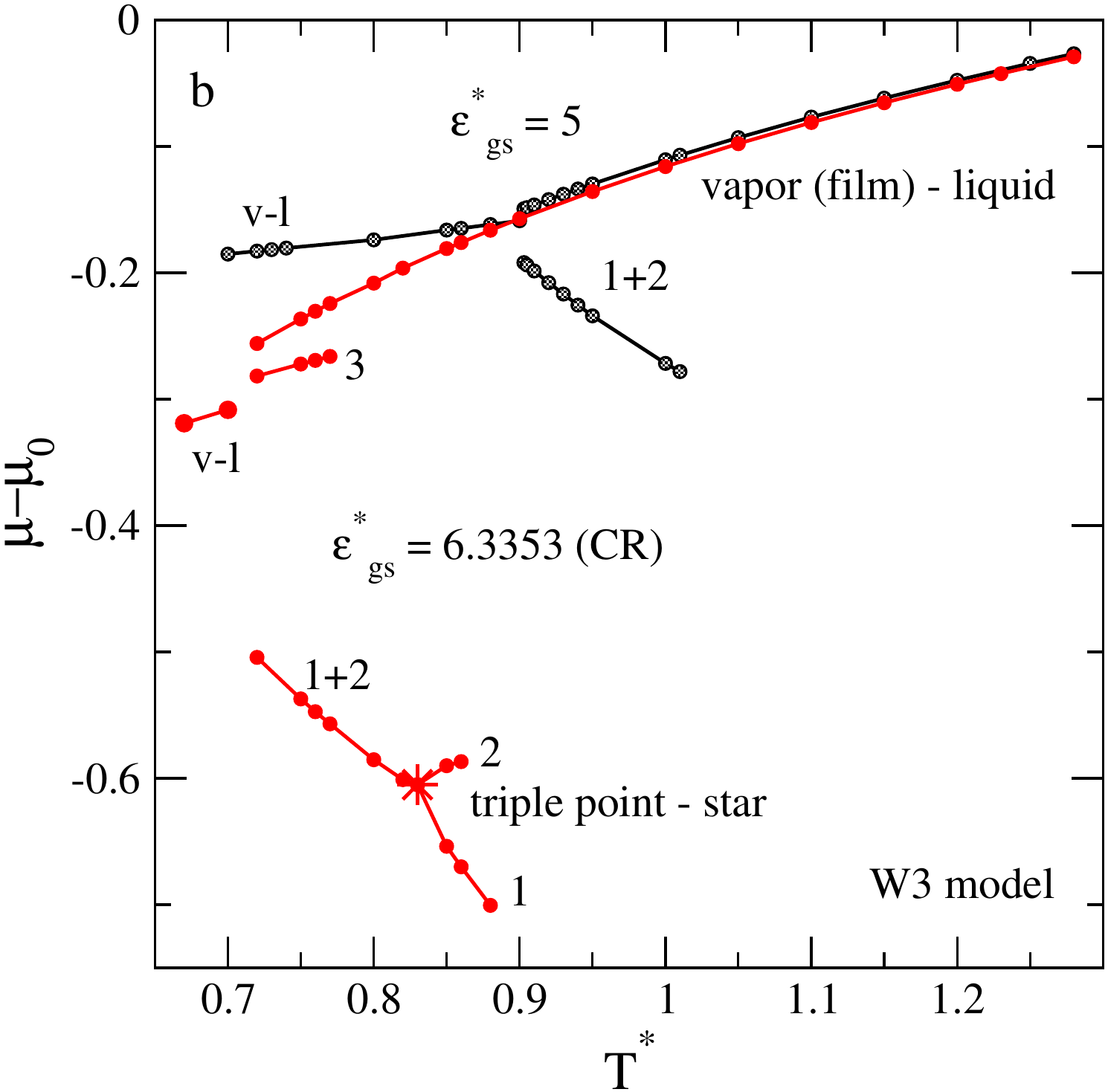}
\end{center}
\caption{(Colour online) Panels a and b: The effect of gas-solid interaction energy on
the density-temperature and chemical potential-temperature projection of the vapor-liquid coexistence envelope for the  W3 model.
The gas-solid energy values are: 5 and 6.3353 (from the combination rule), respectively.
}
\label{fig:10}
\end{figure*}

More pronounced peculiarities of the phase behavior are expected while considering the W3 and W4
models in comparison to the W1 and W2 ones. The principal reason is that the former two models
involve a much narrower square attractive well, cf. table~\ref{tableII}. Immediate consequences can be seen
from the phase diagrams of the W3 model in figure~\ref{fig:10}. Even for a quite weak fluid-substrate
attraction, e.g., $\varepsilon^*_{\text{gs}}=5$, the pore wall is capable of pulling the water molecules and
forming a ``bilayer'' structure. However, this piece of the coexistence is separated from the
principal part. For lower temperatures, water vapor converts into liquid without any peculiarities.
This kind of separation results in a phase diagram consisting of three separate branches as
shown in figure~\ref{fig:10}b. The low temperature branch is just the vapor-liquid coexistence in the
pore, i.e., the capillary condensation. Two branches at higher temperature describe a consecutive
layering ($1+2$ branch) and subsequent condensation that starts from a thin bilayer film. This
thin film is even wider, if one focuses into even higher temperatures. This behavior differs
from the trends observed for W1 and W2 models discussed above.
 
If the strength of water-substrate interaction is chosen from the combination rules, the phase
diagram becomes even more complex. Namely, the pore wall is capable of decoupling the 
$1+2$ ``bilayer'' film into two consecutive layering transitions in the upper part of this
fragment of the diagram with the triple point between them. The critical temperature of the 
first layering transition is higher than of the second one, in contrast to what we have seen just above,
describing the phase behavior of the W2 model. Returning to figure~10a, we observe that at lower temperatures,
the formation of the ``bilayer'' structure via a single transition is followed by one more
transition describing the formation of the third layer. Next, this multilayer structure converts 
into liquid upon increasing the chemical potential. However, the fragments of the phase digram
describing the formation of layers are not connected to the principal envelope, i.e., there are
no triple points. A better illustration of this behavior is shown in figure~\ref{fig:10}b by red lines. The only
triple point is seen between the first and second layering transitions.  

\begin{figure*}[h!]
\begin{center}
\includegraphics[height=6cm,clip]{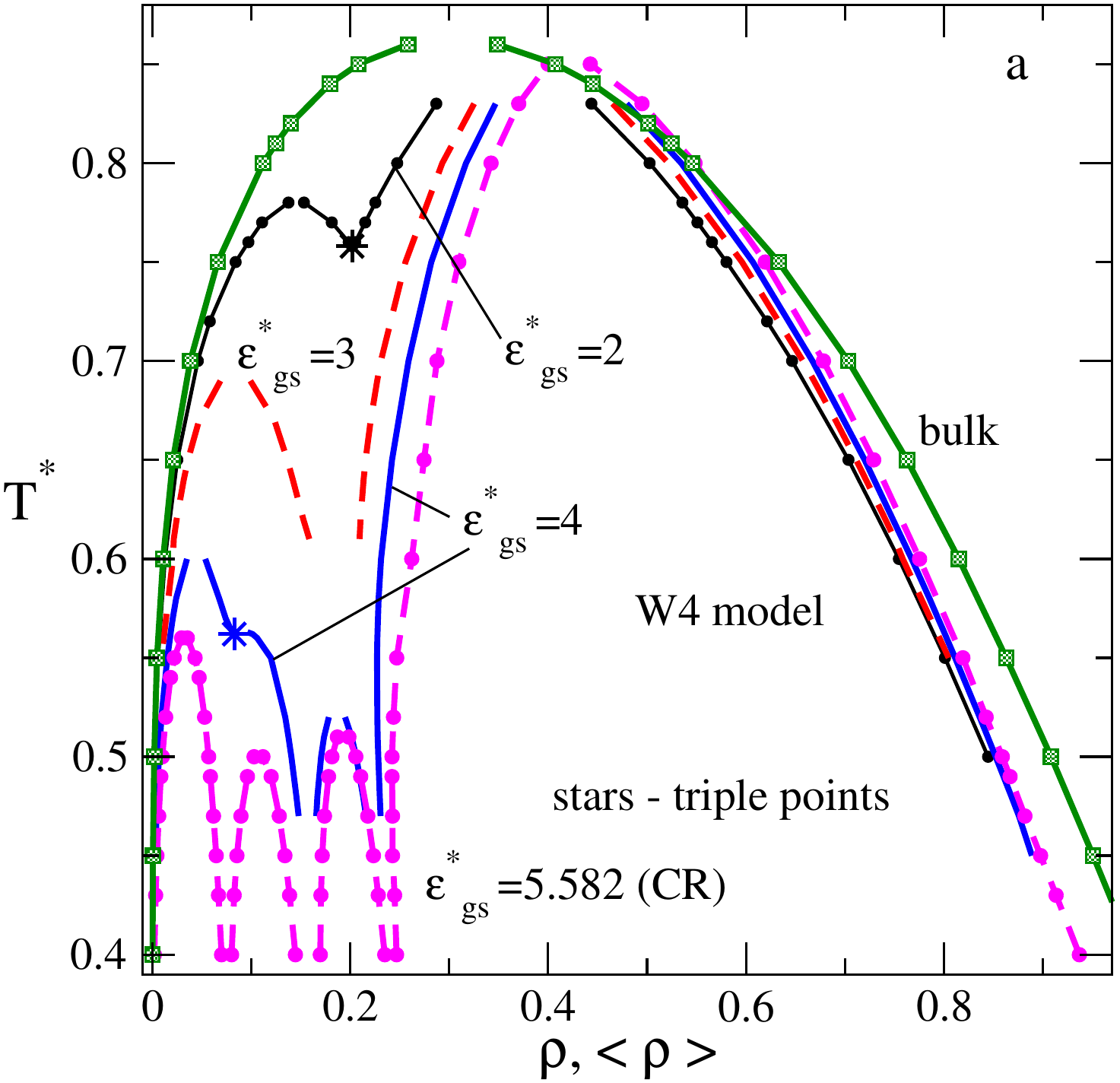}
\includegraphics[height=6cm,clip]{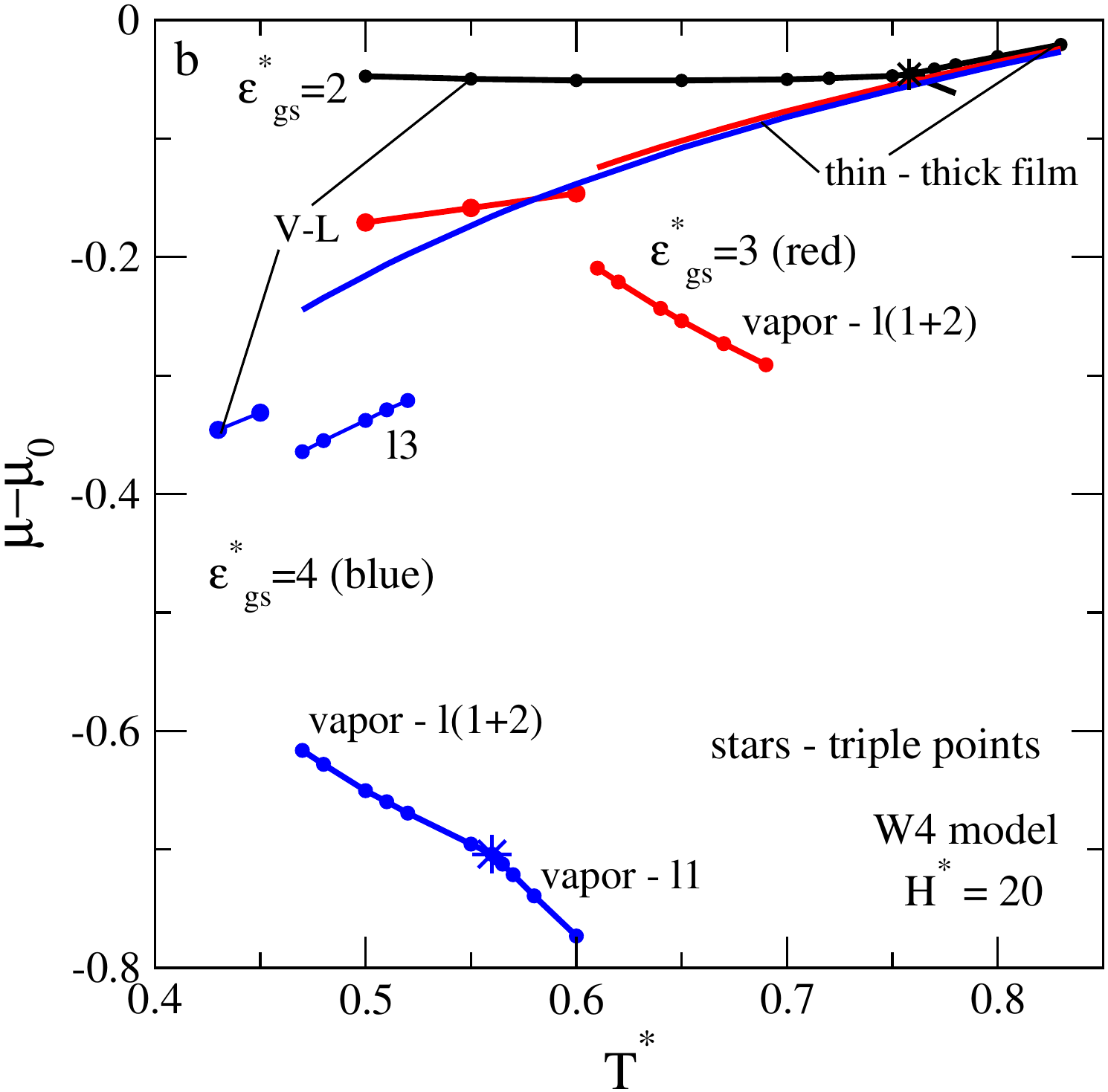}
\end{center}
\caption{(Colour online) Panels a and b: The effect of gas-solid interaction energy on the
density-temperature and
chemical potential-temperature projections of the vapor-liquid coexistence envelope
for the  W4 model.
}
\label{fig:11}
\end{figure*}

More complexity is expected for the W4 model since its parameters correspond 
to the most narrow square well compared to other models. The results of our calculations
illustrate this in two panels of figure~\ref{fig:11}. The phase diagram evolution 
with increasing $\varepsilon^*_{\text{gs}}$
from 2 to 3 keeps track of our discussion while moving from W2 to W3 model. 
In both cases, a ``bilayer'' structure is formed at higher temperatures,
whereas at lower temperatures, capillary condensation occurs. 

Next, the phase diagram corresponding to $\varepsilon^*_{\text{gs}}=4$ exhibits a similarity to the phase
behavior of the W3 model. However, we observe an enhanced difference between the critical
temperatures of the first and second layering transitions that join at the triple point. 
These structures then transform, because of the additional transition due to the formation of 
the third layer. Afterwards, capillary condensation occurs upon increasing the chemical potential
as expected.
The final result in this figure concerns the $\varepsilon^*_{\text{gs}}$ value arising from the 
combination rules. None of the triple points is seen. All three layering transitions are
separated between themselves and separated from the principal condensation envelope.

\begin{figure*}[!t]
\begin{center}
\includegraphics[height=6.5cm,clip]{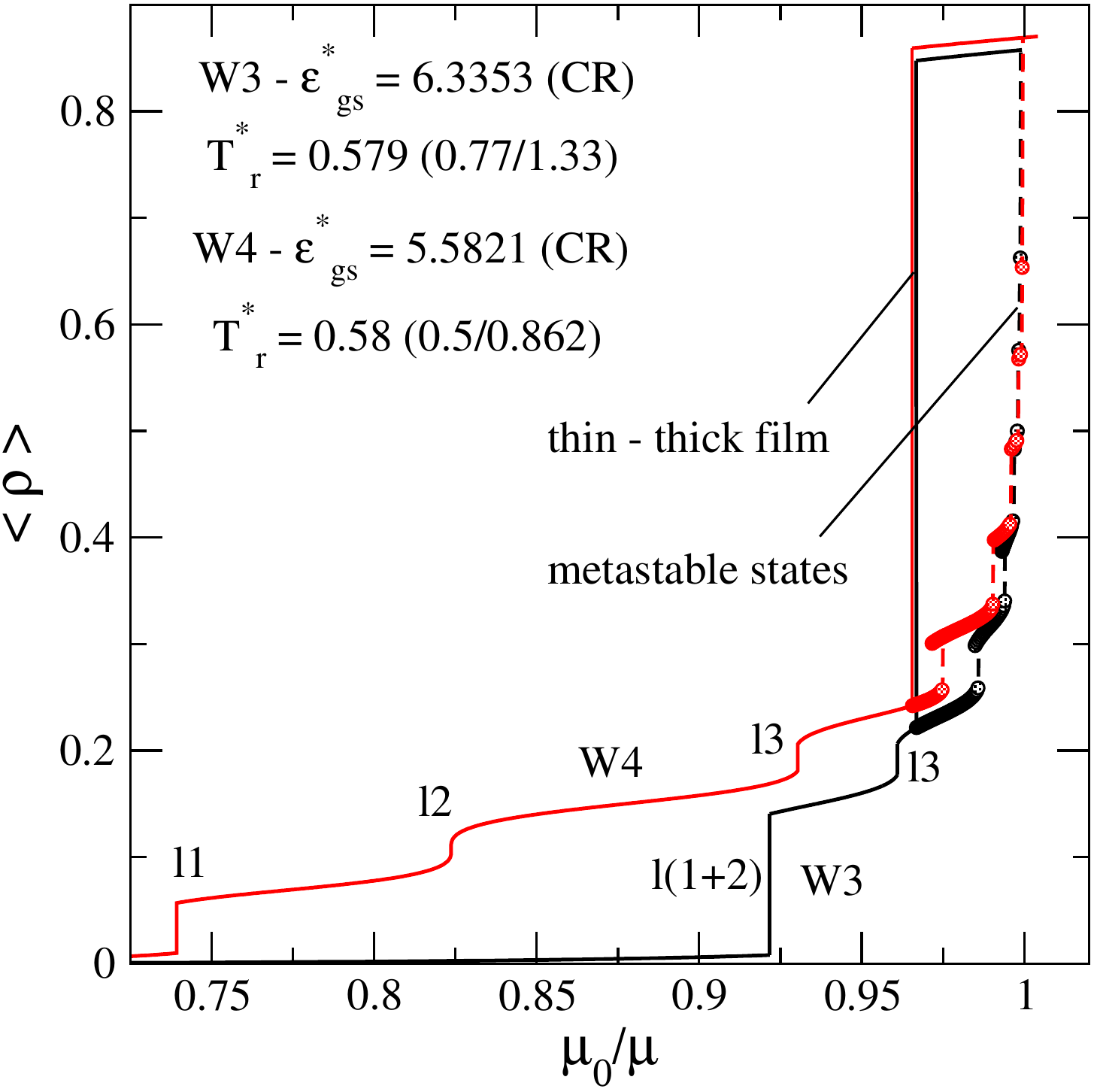}
\end{center}
\caption{(Colour online) Adsorption isotherms of the W3 and W4  water model on
graphite-like surface at similar reduced  temperature. The parameter
$\varepsilon_{\text{gs}}^*$ in both cases follows from the combination rules.
}
\label{fig:12}
\end{figure*}

Additional insight into how the water film grows on the pore walls within these models
with a narrower square well (W3 and W4) comes out from figure~\ref{fig:12}. We compare two examples of the
adsorption isotherms at the same reduced temperature and in both cases consider the
fluid-wall interaction strength value from the combination rules. It can be seen that
the parameter $\lambda$ (square well width) is of much importance. If the attractive square-well is narrower (at a fixed, sufficiently strong association energy),
then the tendency for the formation of several layers on the pore walls is stronger.
The metastable parts of the isotherms consist of consecutive layering transitions. These
metastable parts can be compared with their entirely smooth  ``counterpart'' for
the W1 model (with a much wider square well attraction) in figure~\ref{fig:7ol}a.

\begin{figure*}[!t]
\begin{center}
\includegraphics[height=6.5cm,clip]{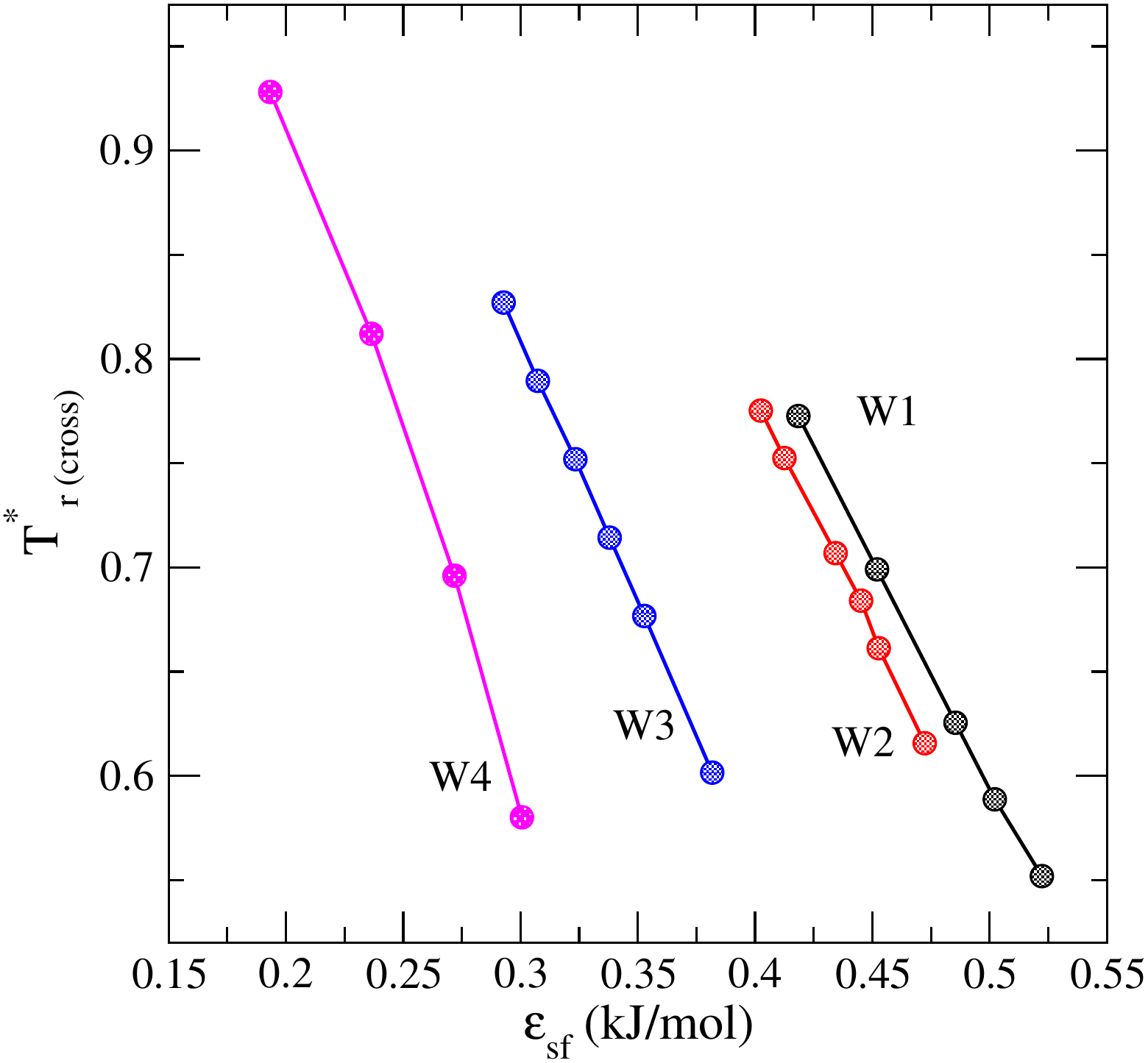}
\end{center}
\caption{(Colour online) Reduced crossover temperature ($T^*_{r(\text{cross})} = T^*_{\text{cross}}/T^*_{c}$)
on the fluid-solid attraction energy $\varepsilon_{\text{sf}} (kJ/\text{mol})$ for all four
water-like models under study. 
}
\label{fig:13}
\end{figure*}

The phase diagrams obtained and discussed above in detail permit to search 
for $\varepsilon^*_{\text{gs}}$ values satisfying the $\mu=\mu_0$ condition at various 
dimensionless temperatures, $T^*$. This search is partially illustrated in figures~\ref{fig:5ol} and \ref{fig:8}.
Next, with $\varepsilon^*_{\text{gs}}$ and $T^*$ available, one can 
obtain the corresponding  $\varepsilon_{\text{sf}}$ values. The crossover temperature as a
function of  water-graphite interaction energy is shown in figure~\ref{fig:13}.

Two observations can be deduced from these curves. Similarity of the behavior of W1 and W2 models is
due to a rather wide square-well attraction of these models. If the balance between the non-associative attraction
and associative site-site bonding is different, like in W3 and W4 models, a shift of the curves to lower
values of $\varepsilon_{\text{sf}}$ is observed. However, the inclination of the curves in the entire set is 
similar. If $\varepsilon_{\text{sf}}$ decreases, the interval of temperature, where one observes a capillary condensation,
shrinks or, in other words, an interval of temperature describing evaporation extends. In
qualitative terms, this behavior can correspond to changes of the wetting temperature.  
One can attempt to evaluate the crossover temperature in pores a few times wider than we 
study at present, in order to approach the description of one-wall phenomena. However, a strict alternative
is to set up the problem with a single wall and permit the growth of the ``macroscopic'' 
film with free terminating interface.

\section{Summary and conclusions}
\label{conclusions}

The water-like models of this study were developed by
Clark et al.~\cite{clark} and describe vapor-liquid coexistence of the bulk water pretty well. 
The inter-particle interaction potential consists of spherically symmetric hard core,
square-well attraction and associative site-site attractive interaction that mimics
hydrogen bonding effects. 
However, we have explored the phase behavior of these models confined to a 
quite wide slit-like pore. To do that, a version of classical density funcional theory is
used. The principal issue we dealt with is in changes of topology of the phase diagrams
upon changing the strength of water-substrate interaction. We considered a hydrophobic pore as well as
moderately hydrophilic and hydrophilic environment for water species. A similar type of investigation 
was performed in \cite{ivan1,ivan2,ivan3,ivan4,ivan5,ivan6,cummings} using
computer simulations techniques. In spite of these efforts, still the problem 
is not comprehensively studied.
A theoretical approach of the present work  has an advantage
over simulations because it intrinsically involves the chemical potential variable 
indispensable in the study of phase equilibria. 

The water-substrate interaction has the form of 10-4-3 potential 
formulated by Steele~\cite{steele} and commonly used in adsorption studies on a graphite surface.
The combination rules were used while deriving values for
energy of interaction between water molecules and graphite. On the other hand,
we assumed various values of the energy to explore in detail the phase behavior of water under confinement.
Our findings qualitatively agree with computer simulation results. In hydrophobic pores, 
interpretation of the behavior of water is much easier than in hydrophilic pores. 
In the latter type of systems, the phase behavior is quite complex and difficult to access by
computer simulation methods. For moderately hydrophilic and hydrophilic setup, we 
established conditions at which additional phases appear at the vapor side of the
coexistence envelope. This is done for all four water-like models in question. Stable
and metastable parts of these rather low density phases were evaluated. 
Triple points under different circumstances were obtained. Continuous and 
discontinuous  growth of the water
film on substrates with different affinity to water was described. In certain
situations, we observed the formation of a bilayer-type water film and the possibility 
of its decomposition into layering transitions. Exploration of the phase behavior
is complemented by the description of adsorption isotherms and of the density distribution
of molecules adjacent to the pore walls. Our results may be combined with
the development of studies of adsorption of water in various pores,
e.g.,~\cite{millan2,malheiro,miqueu}. 

Concerning the missing elements, we would like to mention that the problem of wetting of
this class of models for an associating fluid on graphite requires a separate investigation. 
This should permit to evaluate the wetting temperatures and the contact angles. Important and
interesting extensions should involve exploration of various mixtures in all aspects
of the present work, see e.g.,~\cite{kasia} as a possible starting point.

\section*{Acknowledgements}
 
O.P. is grateful to M. Aguilar for technical support of this work  at the Institute of Chemistry 
of the National University of Mexico (UNAM).
V.M.T. acknowledges the Mexican Ministry of Education (SEP) for the financial support through 
the program PROMEP (M\'{e}xico), UAEH-PTC-831.
\appendix
\section*{Appendix} 
\setcounter{equation}{0}
\def\theequation{A.\arabic{equation}}

In this supplementary material we briefly describe the Density Functional Theory (DFT) that 
was used in the calculations. As it has been already noted, this theory
has been already outlined in detail in \cite{trejos2,c40,c41}. However, the thermodynamic principles
of the approach can be found in \cite{chapter}. Therefore, we refer a reader to the above cited works
for comprehensive explanations.

The system is studied in the Grand Canonical Ensemble. The grand thermodynamic potential, $\Omega[\rho(\mathbf{r})]$, 
and the  Helmholtz free energy, $\tilde A[\rho(\mathbf{r})]$,
are related via the equation,
\begin{equation}
 \Omega[\rho(\mathbf{r})]=\tilde A[\rho(\mathbf{r})]+\int \rd\mathbf{r}[v_f(z)-\mu]\rho(\mathbf{r}),
 \label{A.S1}
\end{equation}
where $\mu$ is the chemical potential of the fluid and $\rho(\mathbf{r})$ is the local density.

We use 
the tilde mark to make a distinction between nonuniform and uniform systems; the latter was described in section~\ref{sec2.2}.
Similarly to the uniform system, the nonuniform system Helmholtz free energy, $\tilde A[\rho(\mathbf{r})]$, 
is divided into the ideal and the
excess parts, $\tilde A[\rho(\mathbf{r})]=\tilde A_{\text{id}}[\rho(\mathbf{r})]+\tilde A_{\text{ex}}[\rho(\mathbf{r})]$.
The ideal part is known exactly,
\begin{equation}
 \tilde A_{\text{id}}[\rho(\mathbf{r})]=kT\int \rd\mathbf{r}\rho(\mathbf{r})\{\ln[\rho(\mathbf{r})\Lambda^3] - 1\},
  \label{A.2}
\end{equation}
while the excess free energy functional 
is considered to be the sum of  the contributions due to short-ranged (hard-sphere) repulsion,  
longer-ranged attractive interactions, and the term resulting from the site-site association,
\begin{equation}
\tilde A_{\text{ex}}[\rho(\mathbf{r})]=\tilde A_{\text{hs}}[\rho(\mathbf{r})]+\tilde A_{\text{att}}[\rho(\mathbf{r})]+\tilde A_{\text{as}}[\rho(\mathbf{r})].
 \label{A.3}
\end{equation}

The hard-sphere contribution, $\tilde A_{\text{hs}}[\rho(\mathbf{r})]$,
is  evaluated in the framework of  the White Bear version of the Fundamental Measure
theory. 
The theory utilizes the concept of four scalar,  $n_{i} (\mathbf{r})$, $i=0,1,2,3$,
and two weighted, ${\bf n}_{i} (\mathbf{r})$, $i=V1,V2$, densities,
that are 
related to the density profile, $\rho (\mathbf{r})$.
This theory and all the  necessary equations were reported
in the work of Yu and Wu \cite{Yu2002}.

The attractive, non-associative part, however,  $\tilde A_{\text{att}}[\rho(\mathbf{r})]$,
is described by using the mean field approximation,
\begin{equation}
 \tilde A_{\text{att}}^{(\text{MF})} =-\frac{kT}{2} \int \rd\mathbf{r} \rd\mathbf{r}' u_{\text{att},ff}(|\mathbf{r}-\mathbf{r}'|)
 \rho(\mathbf{r})\rho(\mathbf{r}'),
  \label{A.4}
 \end{equation}
where $u_{\text{att},ff}(r)$ is given by equation~(\ref{uSW}).

The contribution due to association is given by
\begin{equation}
\tilde A_{\text{as}}= \int \rd\mathbf{r} \Phi_{\text{as}} (\mathbf{r}),
 \label{A.5}
\end{equation}
where $\Phi_{\text{as}}$ is the associative free energy density\cite{trejos2,c40,c41}
that results from a generalization of the first-order thermodynamic perturbation theory (TPT) of Wertheim
to nonuniform systems.
Similarly to the case of the hard-sphere contribution, $\tilde A_{\text{hs}}$,
the evaluation of the  functional $\Phi_{\text{as}} (\mathbf{r})$ requires the
knowledge of the weighted density profiles $n_{i} (\mathbf{r})$
and  ${\bf n}_{i} (\mathbf{r})$~\cite{Yu2002}. 
We have \cite{trejos2,c40,c41}
\begin{equation}
\Phi_{\text{as}}(n_{i}) = 4 n_0(\mathbf{r}) \zeta(\mathbf{r})
\sum_{i=1}^M\left[\text{ln}\chi_i(\mathbf{r})-\frac{1}{2}\left( \chi_i(\mathbf{r})-1\right) \right],  
 \label{A.6}
\end{equation}
where, in contrast to the bulk system, the fraction of non-bonded molecules at a given site $i$, 
depends now on the position vector ${\bf r}$. All the details
and the definition of the quantity  $\zeta(\mathbf{r})$ are given in previous works~\cite{trejos2,c40,c41}.
We only note that the evaluation of the functional $\Phi_{\text{as}}(\mathbf{r})$ requires the knowledge of
the contact value of the pair distribution function. The latter is computed from a generalizations
of the  Carnahan and Starling equation of state  to nonuniform systems~\cite{Yu2002}.   

The density profile results from the minimization of the Grand Canonical potential, cf.~\cite{chapter},
\begin{equation}
 \rho(\mathbf{r})=\rho_b\exp\left\{\frac{1}{kT}\left[\mu_{ex} -V(z)
 - \frac{\delta \left( \tilde A_{\text{hs}}[\rho (\mathbf{r})] + \tilde A_{\text{att}}^{(\text{MF})} + \tilde A_{\text{as}}[\rho (\mathbf{r})]\right) }{\delta \rho(\mathbf{r})} 
\right] \right\},
 \label{A.7}
\end{equation}
where $\rho_b$ is the bulk reference system, i.e., the bulk, uniform system at
the same temperature and the chemical potential as the system under study.
The quantity $\mu_{ex}$ is the configurational chemical potential of the bulk system $\mu_{ex}=\mu-\mu_{\text{id}}$.

The average density of the fluid in  pore, $\left\langle \rho\right\rangle $, 
is calculated from the density profile,
\begin{equation}
 \left\langle \rho\right\rangle = \frac{1}{H} \int_{-H/2}^{H/2} \rd z \rho(z),
 \label{A.8}
 \end{equation} 
while the excess adsorption isotherm for the confined fluid reads,
\begin{equation}
 \Gamma=\int_{-H/2}^{H/2} \rd z [\rho(z)-\rho_b].
 \label{A.9}
 \end{equation}



\newpage
\ukrainianpart

\title[Вода у графітових порах]
{Фазова поведінка водоподібних моделей у нанопорах щілинної форми. Передбачення, які дає теорія функціоналу густини}

\author[О. Пізіо, С. Соколовський, В. М. Трехос]{
	О. Пізіо\refaddr{label1}
	С. Соколовський \refaddr{label2},
	В. М. Трехос\refaddr{label3}
}

\addresses{
	\addr{label1}
Інститут хімії Національного автономного університету Мехіко, Сіркіто Екстеріор, 04510, Мехіко, Мексика
	\addr{label2} 
	Відділ теоретичної хімії університету Марії Склодовської-Кюрі, Люблін 20-031, Польща
	\addr{label3}
Інститут фундаментальних наук та промисловості, Автономний університет штату Ідальго, Карратера Пачука-Тулансінго Км. 4.5, Карбонерас C.P. 42184, Ідальго, Мексика
}


	\makeukrtitle
	\begin{abstract}
		Досліджено фазову поведінку низки водоподібних моделей у щілинних порах наноскопічних розмірів.
		Взаємодія між водою та стінками пор імітує поверхню графіту.
		Певний варіант методу функціоналу густини використовується в якості теоретичного інструменту.
		Моделі води запозичено з роботи Кларка та ін. [Mol. Phys., 2006 \textbf{104}, 3561]. Ці моделі адекватно відтворюють криві співіснування ``рідина-пара'' у об'ємі води. Основний наголос зроблено на змінах топології фазової діаграми води у щілині та встановленню основних тенденцій поведінки температури переходу  між конденсацією та випаровуванням в залежності від потенціалу взаємодії вода-графіт. 
				Ріст водяної плівки на стінках пори проілюстровано у вигляді профілів густини.
		Теоретичні результати обговорюються в контексті даних комп'ютерного моделювання для моделей води у порах.
		\keywords асоціативні флюїди, теорія функціоналу густини, змочування, адсорбція, моделі води
	\end{abstract}
\end{document}